\documentclass[11pt]{article}
\usepackage{geometry}                
\usepackage{graphicx,caption}
\usepackage{amssymb}
\usepackage{nicefrac}
\usepackage{amsmath}
\usepackage{makecell}

\usepackage[usenames,svgnames,table]{xcolor}
\usepackage[colorlinks,allcolors=MediumBlue,breaklinks,hyperfootnotes=true,
	pdftitle={Sneaked references: Fabricated reference metadata distort citation counts},
	pdfauthor={ TBD },
	pdfsubject={Preprint},
	pdfkeywords={sneaked references, undue citations, citation manipulation, metadata registration, bibliometrics, research evaluation}
 ]{hyperref}

\usepackage[natbibapa]{apacite}    
\newcommand{\doi}[1]{\href{https://doi.org/#1}{#1}}
\usepackage{underscore}

\usepackage{comment}
\usepackage{relsize}
\usepackage{nicefrac}
\usepackage{caption}
\usepackage{subcaption}
\usepackage{wrapfig}
\usepackage{fontawesome5}    
\usepackage{halloweenmath}   
\usepackage{tcolorbox} 
\usepackage{hhline}
\usepackage{url}

\newcommand{\SN}{{$\boldsymbol{\mathghost}$ }}
\newcommand{\LO}{{\smaller\faSkullCrossbones}}

\newcommand{\email}[1]{\href{mailto:#1}{#1}}
\newcommand{\mydoi}[1]{\href{https://doi.org/#1}{#1}}

\usepackage{bbding}
\usepackage{subcaption}
\usepackage{array}
\usepackage{siunitx}
\usepackage{booktabs}
\usepackage{comment}
\DeclareMathOperator{\nbrefop}{NbRef}
\newcommand{\nbref}[2]{\nbrefop\left(#1,#2\right)}
\newcommand{\opciten}[3]{#1 \xrightarrow{\;#3\;} #2}
\newcommand{\opcite}[2]{#1 \xrightarrow{\;\;\;} #2}
\newcommand{\opauth}[2]{#1 \text{\hspace{3pt}\raisebox{-2pt}{\NibRight}\hspace{3pt}} #2}
\newcommand{\opjrn}[2]{#1 \text{\hspace{3pt}\faHandshake[regular]\hspace{3pt}} #2}

\title{Detection of metadata manipulations:\\ Finding sneaked references \SN in the scholarly literature}

\author{Lonni Besançon\footnote{
Media and Information Technology, Linköping University, Norrköping, Sweden, 
\email{lonni.besancon@gmail.com},\\
ORCID: \href{https://orcid.org/0000-0002-7207-1276}{0000-0002-7207-1276}
}
\and Guillaume Cabanac\footnote{ Université Toulouse~3 -- Paul Sabatier, IRIT UMR 5505 CNRS, 31062 Toulouse, France; Institut Universitaire de France (IUF), France,
\email{guillaume.cabanac@univ-tlse3.fr},
ORCID: \href{https://orcid.org/0000-0003-3060-6241}{0000-0003-3060-6241}
}
\and Cyril Labbé \footnote{
Univ. Grenoble Alpes, CNRS, Grenoble INP, LIG, 38000 Grenoble, France,
\email{cyril.labbe@univ-grenoble-alpes.fr},
ORCID: \href{https://orcid.org/0000-0003-4855-7038}{0000-0003-4855-7038}
}
\and Alexander Magazinov\footnote{Yandex.Kazakhstan, 43 Dostyq av., Almaty 050010, Kazakhstan,
\email{magazinov-al@yandex.ru},\\
ORCID: \href{https://orcid.org/0000-0002-9406-013X}{0000-0002-9406-013X}
}
\and Jules di Scala\footnote{Université Toulouse~3 -- Paul Sabatier, IRIT UMR 5505 CNRS, 31062 Toulouse, France;
\email{jules.di-scala@univ-tlse3.fr},\\
ORCID: \href{https://orcid.org/0009-0005-3460-0535}{0009-0005-3460-0535}
}
\and
Dominika Tkaczyk\footnote{Crossref, \email{dtkaczyk@crossref.org}, ORCID: \href{https://orcid.org/0000-0001-5055-7876}{0000-0001-5055-7876}
}
\and Kathryn Weber-Boer\footnote{
Digital Science, London, UK, \email{k.weberboer@digital-science.com}, ORCID: \href{https://orcid.org/0000-0002-4495-3001}{0000-0002-4495-3001}
}
}

\date{Started late Summer 2024, version of \today \\
Submitted to \emph{Journal of the Association for Information Science and Technology}}

\begin{document}
\maketitle

\begin{abstract}
We report evidence of a new set of \emph{sneaked references} discovered in the scientific literature.
Sneaked references are references registered in the metadata of publications without being listed in reference section or in the full text of the actual publications where they ought to be found.
We document here $80,205$ references sneaked in metadata of the \emph{International Journal of Innovative Science and Research Technology (IJISRT)}.
These sneaked references are registered with Crossref and all cite---thus benefit---this same journal.
Using this dataset, we evaluate three different methods to automatically identify sneaked references.
These methods compare reference lists registered with Crossref against the full text or the reference lists extracted from PDF files. 
In addition, we report attempts to scale the search for sneaked references to the scholarly literature.
\end{abstract}

\section{Introduction}
\label{sec:intro}
Citation-based indices or metrics like the $h$-index~\citep{HIndex}, the Journal Impact Factor \citep{Garfield1994} or the Field-Weighted Citation Impact (FWCI) \citep{PURKAYASTHA2019635} are cornerstones to many rankings: Clarivate's ‘Highly Cited Researchers’ list\footnote{\url{https://clarivate.com/highly-cited-researchers/}}, the Shanghai Ranking\footnote{\url{http://www.shanghairanking.com/}}, Times Higher Education World University Rankings\footnote{\url{https://www.timeshighereducation.com/world-university-rankings}}, QS World University Rankings\footnote{\url{https://www.topuniversities.com/qs-world-university-rankings}}, or U.S. News Education Rankings\footnote{\url{https://www.usnews.com/best-colleges/rankings/}}.
These citation-based performance metrics are provided by various scientometrics services: \href{http://scholar.google.com}{Google Scholar} ($h$-index), \href{https://openalex.org}{OpenAlex} ($h$-index), \href{https://www.scopus.com}{Scopus} ($h$-index and FWCI), and the \href{https://www.webofscience.com}{Web of Science} ($h$-index and Journal Impact Factor); \href{http://www.dimensions.ai}{Dimensions} provides the Field Citation Ratio (FCR) and Relative Citation Ratio (RCR)\citep{BodeEtAl2023}.

Practically speaking, the computation of these indicators requires processing of the metadata describing scientific publications: authors, institutions, reference lists, registration dates, attributing fields to journals and publications, and--critical to the research presented here--reference lists.
Crossref\footnote{\url{https://www.crossref.org/}} provides infrastructure for registering metadata for scholarly works, including a DOI. To use this infrastructure, organisations join Crossref as members. In many cases, the metadata registered by Crossref members include reference lists. The identifiers of cited works are either provided by Crossref members or automatically added where matching is possible. Crossref is one of the major sources of scholarly data for publishers, authors, librarians, funders, and researchers \citep{HendricksEtAl2020}.
Various scientometrics services like Dimensions, OpenAlex, or SpringerLink make use of metadata deposited with Crossref.

Citation gaming to artificially boost citation-based metrics occurs in various forms~\citep{biagioli2020gaming}.
While most of them involve simply adding references to the research papers directly through a varied set of methods and actors \citep[see, e.g.][]{BeelAndGipp2010,Davis2016,FoleyAndValkonen2012,Franck1999,HeathersAndGrimes2022,Kojaku2021,Labbe2010}, \emph{sneaked references} offer a different pathway to citation gaming \citep{Besan2024Sneaked}.
The underlying strategy behind \emph{sneaked references} is to inject irrelevant and undue citations into the metadata of an accepted article at the time of its registration with scientific repositories. 
Sneaked references are only present in the metadata of the article and are not part of the actual reference list of this document where they should be found. 
This malpractice generates undue citations that artificially inflate citation counts.

\begin{figure}[htbp]
\begin{center}
\includegraphics[width=.9\textwidth]{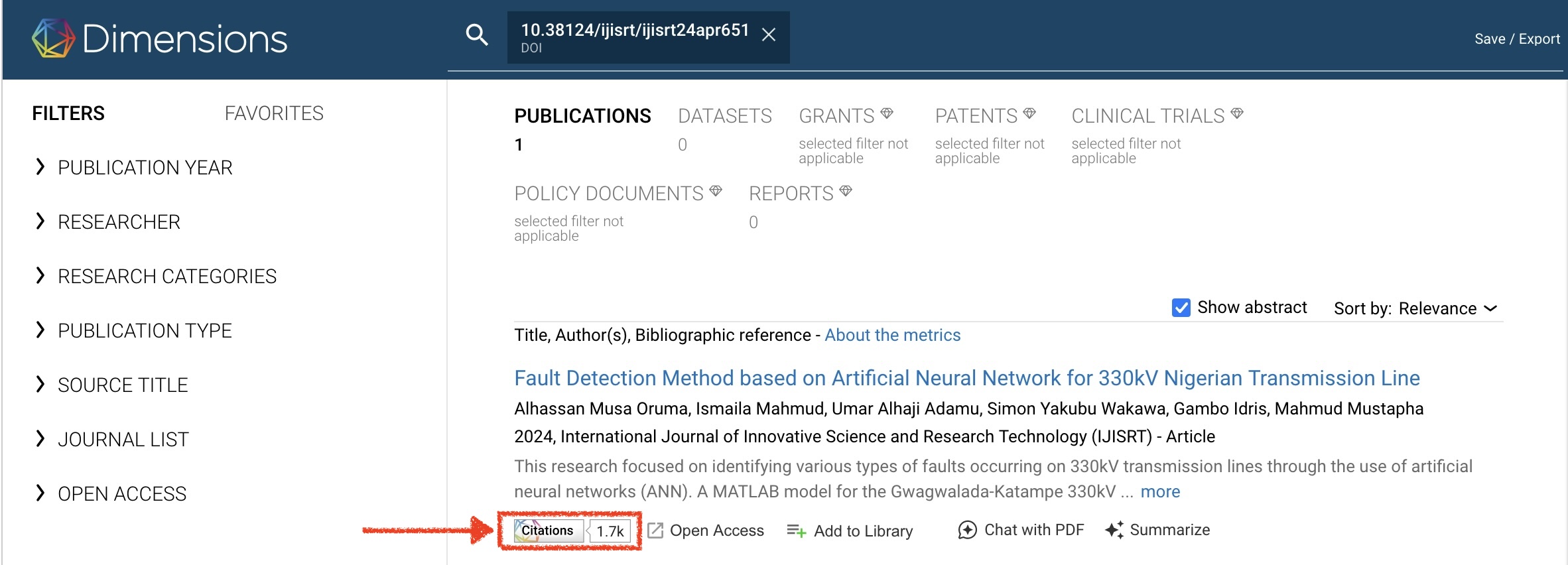}
 \captionsetup{width=.8\linewidth}
\caption{
The citation count of \doi{10.38124/ijisrt/ijisrt24apr651} is 1.7k according to \href{https://app.dimensions.ai/discover/publication?search_mode=content&search_text=10.38124\%2Fijisrt\%2Fijisrt24apr2251&search_type=kws&search_field=doi}{Dimensions}: Early Dec. 2024 it benefits from at least $6,059$ sneaked references (see \autoref{fig:BenfBarChart}). There is no reason to think that authors are responsible for this discrepancy.}
\label{Fig:Dim}
\end{center}
\end{figure}

\begin{figure}[htbp]
\begin{center}
\includegraphics[width=.9\textwidth]{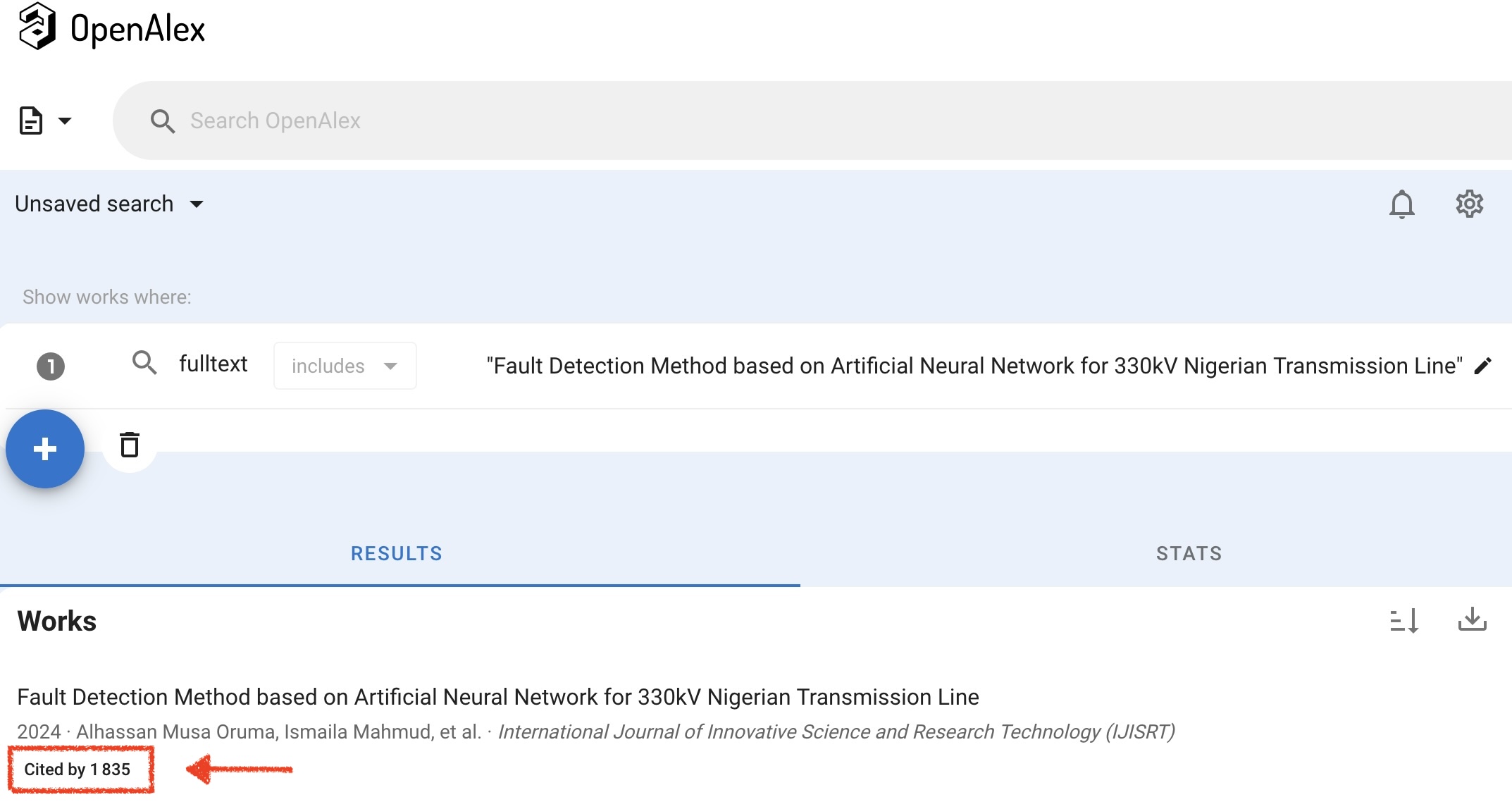}
 \captionsetup{width=.8\linewidth}
\caption{
The citation count of \doi{10.38124/ijisrt/ijisrt24apr651} is 1.8k according to \href{https://openalex.org/works/w4396228980}{OpenAlex}: Early Dec. 2024 it benefits from at least $6,059$ sneaked references (see \autoref{fig:BenfBarChart}). 
There is no reason to think that authors are responsible for this discrepancy.}
\label{Fig:OpenAlex}
\end{center}
\end{figure}

In this article, we report $2,782$ Crossref records spoiled with at least $80,205$ references sneaked into their metadata reference lists. All sneaked references benefit the same journal, namely, the journal in which the reference lists were published.
The paper benefiting the most from sneaked references received a total of $6,059$ undue citation counts, some of which did make their way into various scientometrics services (see \autoref{Fig:Dim} and~\ref{Fig:OpenAlex}).

We designed and evaluated two different methods to automatically identify sneaked references by comparing references registered with Crossref against either the raw text or the reference lists extracted from PDF files. Both methods assume that references registered with Crossref are registered with enough information that they can be found in the extracted text (e.g., {\tt unstructured} attribute). 

The first method $\mathcal M_1$ identifies in the list registered with Crossref.
This method depends on an identical order of elements in the two lists and assumes that sneaked references appear at the end of the list registered with Crossref. 

The second method $\mathcal M_2$ automatically identifies each and every reference registered with Crossref in the raw text extracted from the PDF file. 
The rationale behind this method is that a particular reference field in a Crossref record reflects closely the text of this reference in the PDF file.

These two new methods are compared to an existing approach, $\mathcal M_0$ presented in \citep{Besan2024Sneaked}, which relies on a comparison of reference lists lengths. This approach was found to be effective in providing a lower bound to the number of sneaked references, by comparing reference lists retrieved from HTML document versions to the reference lists registered with Crossref.

These methods work only at the document level.
To identify sneaked references in the scientific literature as a whole, one of these methods must be applied to each and every document, individually.  
We report here the result of an attempt to identify sneaked references at a large scale by applying method $\mathcal M_0$ on 47,170,721 documents previously processed by Dimensions, published since the year 2000.
For each of these documents, the reference list was extracted from the PDF file and stored in a database to be compared with metadata registered with Crossref.

Previous work \citep{Besan2024Sneaked} has mentioned that in data registered with Crossref, duplicated references sometimes appear together with sneaked references. We therefore attempted to identify duplicated references in Crossref metadata in the hopes of identifying new cases of sneaked references. 
\autoref{sec:method} presents the dataset and explains in detail $\mathcal M_1$ and $\mathcal M_2$.
The proposed methods are assessed using the collected dataset. 
\autoref{sec:results} gives precise information about the $80,205$ references sneaked in metadata of the \emph{International Journal of Innovative Science and Research Technology (IJISRT)}: when and where they were sneaked in, to the benefit of which document, and so on.
\autoref{sec:Large} provides insight from attempts to detect sneaked references at a larger scale, including the systematic challenges that inhibit these efforts.
\autoref{sec:conclu} concludes with a discussion of some of the known routes to erroneous references, the actors involved, and some recommended actions which could address the problem of sneaked references.

\section{Dataset and comparison Methods}
\label{sec:method}

\autoref{sec:data} presents the information upon which the dataset was identified and details about how it was retrieved.
Sections~\ref{sec:meth1} and \ref{sec:meth2} give a detailed descriptions of the proposed methods ($\mathcal M_1$, $\mathcal M_2$ respectively).
Sections~\ref{perf:M1},~\ref{perf:Comp} and ~\ref{perf:M0} provide performances results.

\subsection{The \emph{International Journal of Innovative Science and Research Technology (IJISRT)}}
\label{sec:data}

Visual inspection of several PDF files of the \emph{International Journal of Innovative Science and Research Technology (IJISRT)}, and their corresponding Crossref {\tt json} records reveals that, in some cases, references are sneaked in at the end of the Crossref {\tt reference-list} attribute.

On 24 July 2024, Cristian Consonni alerted some of the authors about an entry of the \href{https://www.irit.fr/~Guillaume.Cabanac/problematic-paper-screener}{Problematic Paper Screener} \citep{cabanac:hal-03829578} that highlighted tortured phrases in a certain \emph{IJISRT} article (DOI: \doi{10.38124/ijisrt/ijisrt24apr2410} -- \href{https://pubpeer.com/publications/12E6CBB622A5CA78D44ED95192A0C7}{PubPeer}). 
He further noted that the article had 237 citations, which was unusually high for an article published in April 2024, only 3 months earlier.
Further inspection revealed that there were a significant number of other articles in \emph{IJISRT} with a seemingly disproportionate number of citations and that most---if not all---citations had come from the same journal. As a result of this discovery, we sought citations in the actual text of the citing articles in vain, which indicates a pattern of sneaked references.

On the same day (24 July 2024), we queried \href{https://app.dimensions.ai}{Dimensions} for all articles in \emph{IJISRT} and retrieved the resulting CSV file, including the list of corresponding DOIs. 
For each retrieved DOI, the corresponding PDF file was downloaded from the publisher website (29 August 2024). 
Additionally, for all DOIs, Crossref records were downloaded from Crossref using the relevant API (28--29 August 2024).
This served as a development dataset, and on 25 November 2024 we downloaded the final dataset presented here.

In this final dataset, the observed sneaked references are always benefiting papers with DOIs prefixed with \href{https://www.ijisrt.com}{10.38124/ijisrt}.
This prefix identifies the \emph{International Journal of Innovative Science and Research Technology (IJISRT)}. 
All observed sneaked references appear in Crossref records after the expected references, as an irrelevant addendum to the reference list.
We used these two properties (journal-level self-citation and position at the bottom of the sneaked references in Crossref metadata, {\tt reference-list} attribute) to study to which extent sneaked references occur.

\subsection{$\mathcal M_1$: Comparing Crossref records with references extracted from PDFs}
\label{sec:meth1} 

The idea is to extract a reference list from the PDF files for them to be compared with the ones registered with Crossref.
Extracting the reference list from a PDF file can be done using a tool that transforms PDF files into XML files. 
In XML format, the reference list is clearly identified and can be automatically analysed.
 
\begin{figure}[h!]
    \centering 
      \includegraphics[width=.9\linewidth]{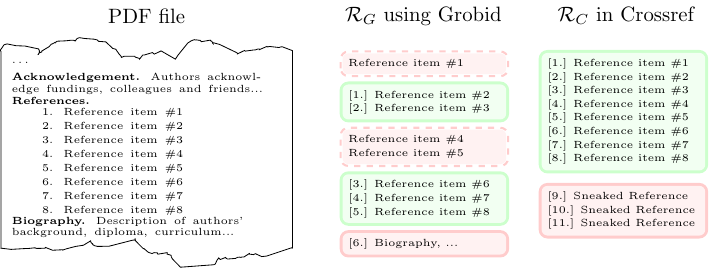}
       \captionsetup{width=.8\linewidth}
    \caption[0.8\linewidth]{A PDF file with a list of 8 references. The reference list extracted by Grobid ($\mathcal R_{G}$) does not contain some of the expected references (e.g., references \#1, \#4, and \#5) and does feature non-existing references (e.g., references $\mathcal R_{G}$ [6.]). The reference list registered with Crossref ($\mathcal R_{C}$) contains 3 sneaked references: [9.], [10.], [11.]. This is a {\bf Case~3} situation.}\label{fig:lists}
\end{figure}

Our process was the following for each collected DOI:
\begin{itemize}
    \item The reference list $\mathcal R_{C}$ registered with Crossref is built from the json file provided by Crossref.
	In the following, $Last_{C}$ denotes the last element of the list $\mathcal R_{C}$.
 
    \item A reference list, $\mathcal R_{G}$ (in XML format) is extracted from the PDF file using Grobid~\citep{GROBID} (default configuration). 
	Unfortunately, Grobid, while being quite reliable, sometimes skips some references. 
	This results in missing references in $\mathcal R_{G}$. 
	Items might be missing in bulk, either at the beginning, end or middle of the reference list.
	In very exceptional cases, Grobid inserts {\it hallucinated} references: $\mathcal R_{G}$ might contain references that do not appear in the PDF. 
	We spotted cases where Grobid added the biography of an author as the last item of $\mathcal R_{G}$.
	This can happen when the bibliography appears, in the PDF, just after the last entry of the reference section (see the left panel in \autoref{fig:lists}).
	Nevertheless, we'll use the last reference of $\mathcal R_{G}$, which we denote $Last_{G}$.
\end{itemize}

Comparing $Last_{G}$ to $Last_{C}$ gives information about both sneaked references and Grobid's capability to identify correctly the last reference in the PDF's reference list.
Let us consider the three following cases:
\begin{itemize}\setlength{\itemindent}{2em}
    \item[\bf Case 1.] If $Last_{C}=Last_{G}$ we conclude that $\mathcal R_{C}$ is correct with regards to the PDF version.
    The Crossref record does not contain any sneaked references. 
    
    \item[\bf Case 2.]\label{Case2} If $\exists r \in \mathcal R_{C}$ such that $r=Last_{G} \land r \neq Last_{C}$ we conclude that references appearing after $r$ in $\mathcal R_{C}$ are sneaked references, forming a list denoted $\mathcal L_{\boldsymbol{\mathghost}}$.
	In the specific dataset of \href{https://www.ijisrt.com}{10.38124/ijisrt}, close analysis of $\mathcal L_{\boldsymbol{\mathghost}}$ instances reveal that, from time to time, the first items are not sneaked references.
	This happens when Grobid omits to extract from the PDF the end of the reference section, resulting in a truncated $\mathcal R_{G}$.
	Considering that in all inspected cases, all the sneaked references concerned DOIs starting with $10.38124$, we decided to remove from $\mathcal L_{\boldsymbol{\mathghost}}$ all elements preceding the first appearance of a DOI starting with $10.38124$.
	In other words, we skipped the references at the top of the list when they are not prefixed by $10.38124$.
    Without this {\bf cleaning operation}, some of the legitimate references would have been wrongly considered as sneaked references.  

    \item[\bf Case 3.] If $\nexists r \in \mathcal R_{C}$ such that $r=Last_{G}$ we can conclude that $Last_{G}$ is an artifact created by Grobid.
    $Last_{G}$ is a hallucinated reference that does not appear, neither in $\mathcal R_{C}$ nor in the PDF file. 
    In that case no conclusion can be drawn.
	Nevertheless, in the specific dataset of \href{https://www.ijisrt.com}{10.38124/ijisrt}, a close analysis reveals that---sometimes---sneaked references can be found at the end of $\mathcal R_{C}$.
	Again, in all inspected cases, the sneaked references are benefiting DOIs starting with $10.38124$.
	We decided to build $\mathcal L_{\boldsymbol{\mathghost}}$ with all elements of $\mathcal R_{C}$ appearing after the last element that has not a DOI starting with $10.38124$. 
	Without this {\bf backward check} (iterating over the list from the bottom to the first non $10.38124$ DOI), some of the sneaked references would have been wrongly considered as sneaked references.
\end{itemize}

\autoref{fig:lists} illustrates a {\bf Case~3} situation. 
For a DOI like this, the list $\mathcal L_{\boldsymbol{\mathghost}}$ is built from the trailing references with DOIs starting with $10.38124$.
This method could be used for every DOI.
Identifying {\bf Case~1} and {\bf Case~2} is a way to evaluate how precise the extraction of sneaked references using Grobid can be.  

\subsection{$\mathcal M_2$: Comparing Crossref records with the full text extracted form PDFs}
\label{sec:meth2} 

The rationale behind this method is that the field {\tt unstructured} of a particular reference entry in a Crossref record reflects closely the text of this reference in the PDF file.
As a consequence, the character string $s$ found at this field must appear in the text $\mathcal T$ extracted from the PDF file.
There is even no need to restrict the search of $s$ to the reference section. 
As shown in~\autoref{sec:meth1}, identifying correctly the reference section is challenging.
Unstructured references are quite long, thus the search of $s$ in $\mathcal T$ is unlikely to generate false positive.

More specifically, the following steps were performed for every DOI:

\begin{enumerate}
    \item The reference list $\mathcal R_{C}$ registered with Crossref is built from the json file provided by Crossref.
            In the following, $s$ denotes an element of the list $\mathcal R_{C}$.

    \item The full text $\mathcal T$ is extracted from the PDF file using the \href{https://pypi.org/project/pypdf/}{pypdf Python library}.

    \item $\forall s\in {\mathcal R_{C}}$, a search of $s$ in $\mathcal T$ is performed.
            The goal is to identify $s'$
            the substring of $\mathcal T$ that is the closest to $s$ according to $\delta(s,s')$ the \href{https://rapidfuzz.github.io/Levenshtein/levenshtein.html#ratio}{normalized Levenstein distance}\footnote{\textit{partial_ratio} method from the \href{https://pypi.org/project/RapidFuzz/}{RapidFuzz Python library}}.
            The similarity $\delta(s,s')$ is ranging from 0 (totally different) and 100 (entirely similar).

    \item If $\delta(s,s')<60$ this means that no character string $s'$ highly similar to $s$ could be found in $\mathcal T$. 
            This most probably happens when the reference $s\in \mathcal R_{C}$ does not exist in the document, revealing that $s$ is a sneaked reference. 
            On the contrary, if $\delta(s,s') \geq 60$ a character string $s'$ quite similar to $s$ exists in $\mathcal T$.
            In that case $s$ is not a sneaked reference.
            The $60$ threshold was set experimentally, after manual examination of several cases.
\end{enumerate}

\subsection{Measuring the performance of $\mathcal M_1$ the detection method using the ‘last’ element of reference lists}
\label{perf:M1}

We consider here the $3,132$ records with $\mathcal R_{C}\neq\emptyset$ and $\mathcal R_{G}\neq\emptyset$ that contain a total of $78,736$ sneaked references. 
These records are distributed among the three identified cases (see~\ref{sec:method}) as follows: 

\begin{itemize}
\setlength{\itemindent}{2em}
    \item[\bf Case 1.] $331$ DOIs ($10.5\%=\nicefrac{331}{3,132}$) with no sneaked references where correctly identified because $Last_{C}=Last_{G}$.
    
    \item[\bf Case 2.] When $\exists r \in \mathcal R_{C}$ such that $r=Last_{G} \land r \neq Last_{C}$ then $\mathcal L_{\boldsymbol{\mathghost}}$ is composed of references appearing after $r$ in $\mathcal R_{C}$. A cleaning operation (\autoref{Case2}) might be needed. 
    \begin{itemize}
        \item No Cleaning needed: For $1,788$ ($ \%=\nicefrac{1,788}{3,132}$) DOIs, $\mathcal L_{\boldsymbol{\mathghost}}$ was correct. This represents a total of $46,297$ ($58.8\%=\nicefrac{46,297}{78,736}$) sneaked references.
        
        \item Cleaning needed: For $840$ DOIs ($ \%=\nicefrac{840}{3,132}$), $\mathcal L_{\boldsymbol{\mathghost}}$ contains potential false positives: references that would have been classified as sneaked without the cleaning operation. This represents a total of $2032$ references ($2.2\%=\nicefrac{2032}{78,736}$).
    \end{itemize} 
    
    \item[\bf Case 3.] For $173$ DOIs, $\nexists r \in \mathcal R_{C}$ such that $r=Last_{G}$. A backward check (\autoref{Case2}) might be needed to identify sneaked references. For the $173$ instances of this case, the backward check identifies sneaked references. Without this check $3,176$ sneaked references would have been undetected ($4\%=\nicefrac{3176}{78,736}$).
\end{itemize}

\subsection{Comparing $\mathcal M_1 $ and $\mathcal M_2$}
\label{perf:Comp}

Method $\mathcal M_1 $ relies to a high extent on the tool use to extract the reference list from the PDF. 
We did implement this method using Grobid.
Performances of the method is thus dependant of the ability of Grobid to accurately extract the reference list. 
Since this task is not trivial, and Grobid occasionally makes mistakes, we had to use additional assumptions about the sneaked references to get reliable results: sneaked references appear after the last genuine reference extracted from the PDF file.
This means that this method might not generalize easily to other instances of the sneaked references problem.

Method $\mathcal M_2$ is not dependant on identifying the reference list in the PDF file and does not make any assumptions about where sneaked references are.
As such, this method should generalize better than the first one. 
Nevertheless, in some corner cases in the text extracted from the PDF, additional text fragments like headers or footers might appear in the middle of a reference, thus making the reference identification impossible.

One common drawback of both methods is that the field {\tt unstructured} must be correctly deposited with Crossref by publisher for the methods to work properly: one could imagine cases where only reference DOIs are provided.

For every DOI with references in the metadata and available PDF, we compared the numbers of sneaked references reported by both methods (see~\autoref{tab:comp}). 
For this dataset, among $3,186 (= 2,953 + 233)$ compared DOIs, the methods disagreed in 233 (7.3\%) cases.
Among these, for only 11 DOIs a difference greater than 10 is observed for the number of sneaked references reported.
This explains why the total numbers of sneaked references detected by the methods differ only by $0.9\%$ ($\nicefrac{80,909-80,205}{80,205}$). 

It seems that most discrepancies are due to cases where the first method underestimated the number of sneaked references.

\begin{table}[!htb]
    \begin{subtable}{.5\linewidth}
      \centering
        \begin{tabular}{|lr|}
            \hline
            Total processed DOIs & 4,077 \\
            \hline
            -- DOIs with no references in JSON & 855 \\
            -- DOIs with no PDF & 36 \\
            -- DOIs where methods agreed & 2,953 \\
            -- DOIs where methods disagreed & 233 \\
            \hline
        \end{tabular}
        \label{tab:a}
        \caption{}
    \end{subtable}%
    \begin{subtable}{.5\linewidth}
      \centering
        \begin{tabular}{|l|r|r|}
         \hline
            Method & DOIs manipulated & sneaked references \\
         \hline
            $\mathcal M_1$ & 2,782 & 80,205 \\
         \hline 
            $\mathcal M_2$ & 2,787 & 80,909 \\
        \hline
        \end{tabular}
        \label{b}
        \caption{}
    \end{subtable} 
    \caption{Statistics on DOIs (a) and comparison of methods findings (b)}
    \label{tab:comp}
\end{table}

\subsection{Measuring the performance of $\mathcal M_0$ that uses the lengths of registered and extracted reference lists}
\label{perf:M0}

Comparing the list lengths, (adapting \citep{Besan2024Sneaked} that uses HTML reference lists and Crossref reference lists) would give $84,270$ sneaked references.
This is an overestimation of $5,534$ ($7\%=\nicefrac{(84,270-78,736)}{78,736}$) of the total number of sneaked references.

For some DOIs the error can be quite high (max = 465).
Relying solely on lists length comparison will generates many false positive.

This an important drawback of method $\mathcal M_0$ when implemented using Grobid. 
It suggests that reference lists extracted using Grobid tend to be shorter than the ones actually existing in documents.

Using a length comparison method will thus overestimate the number of sneaked references.  
Using either the last element method $\mathcal M_1$ or the raw comparison method $\mathcal M_2$ is a more accurate way detect sneaked references.

The next section provides a detailed description of the results obtained using the systematic comparison of Grobid output vs Crossref records ($\mathcal M_1$).

\section{Characteristics of the sneaked references found in \emph{IJISRT} }
\label{sec:results}

We review the characteristics related to the \emph{IJISRT} dataset: When and where the sneaked references were inserted and to whom they benefit? 

Detailed examination of sneaked references properties aims at delineating the source and the nature of their existences.

\subsection{Broad overview: How many? Where and when references were sneaked in? Who are the beneficiaries?}

\begin{figure}[htbp]
\centering
	\begin{subfigure}{.45\textwidth}
	  	\centering
		\includegraphics[width=\linewidth]{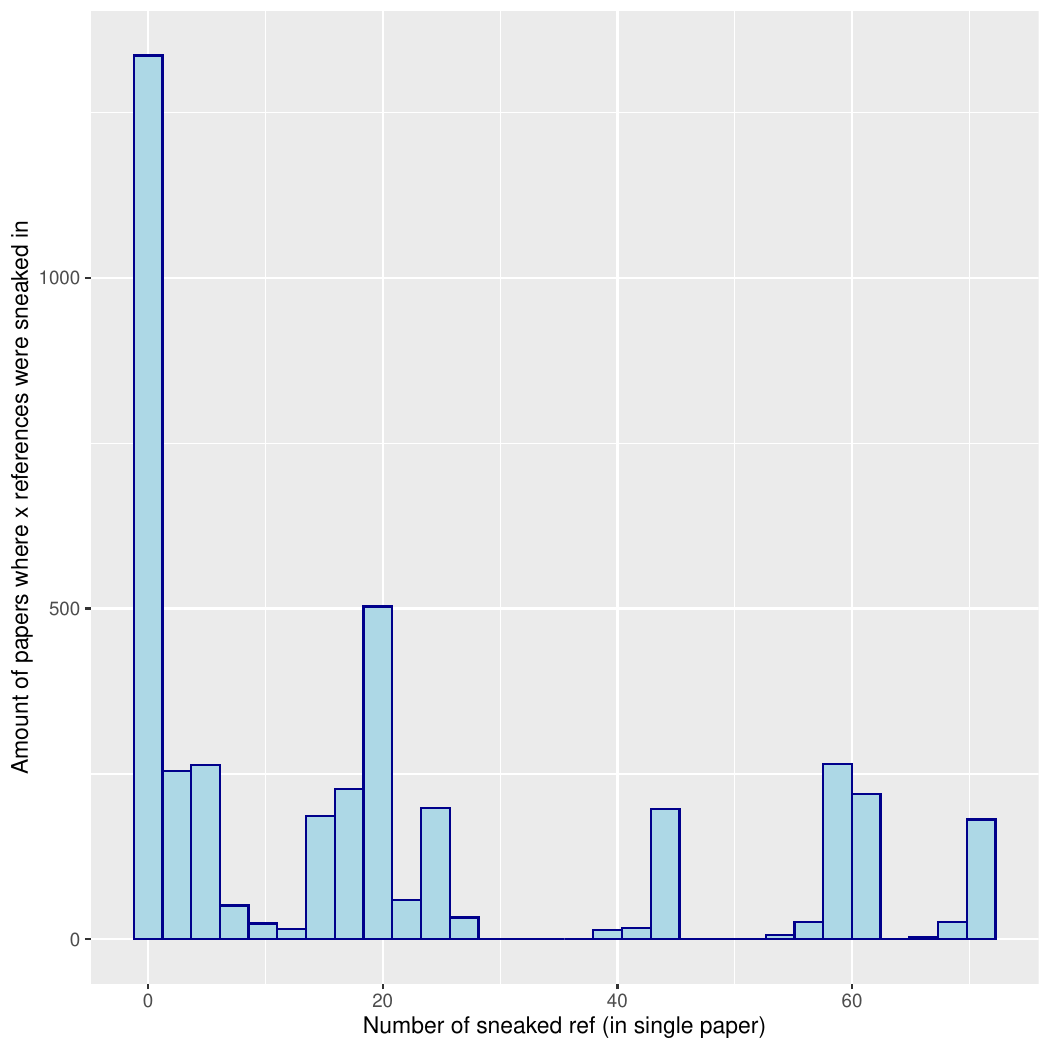}
		\caption{All DOIs, including those with zero sneaked references. 
		}
		\label{Fig:SneakedPerPaperBar}
	\end{subfigure}
	\begin{subfigure}{.45\textwidth}
	  	\centering
		\includegraphics[width=\linewidth]{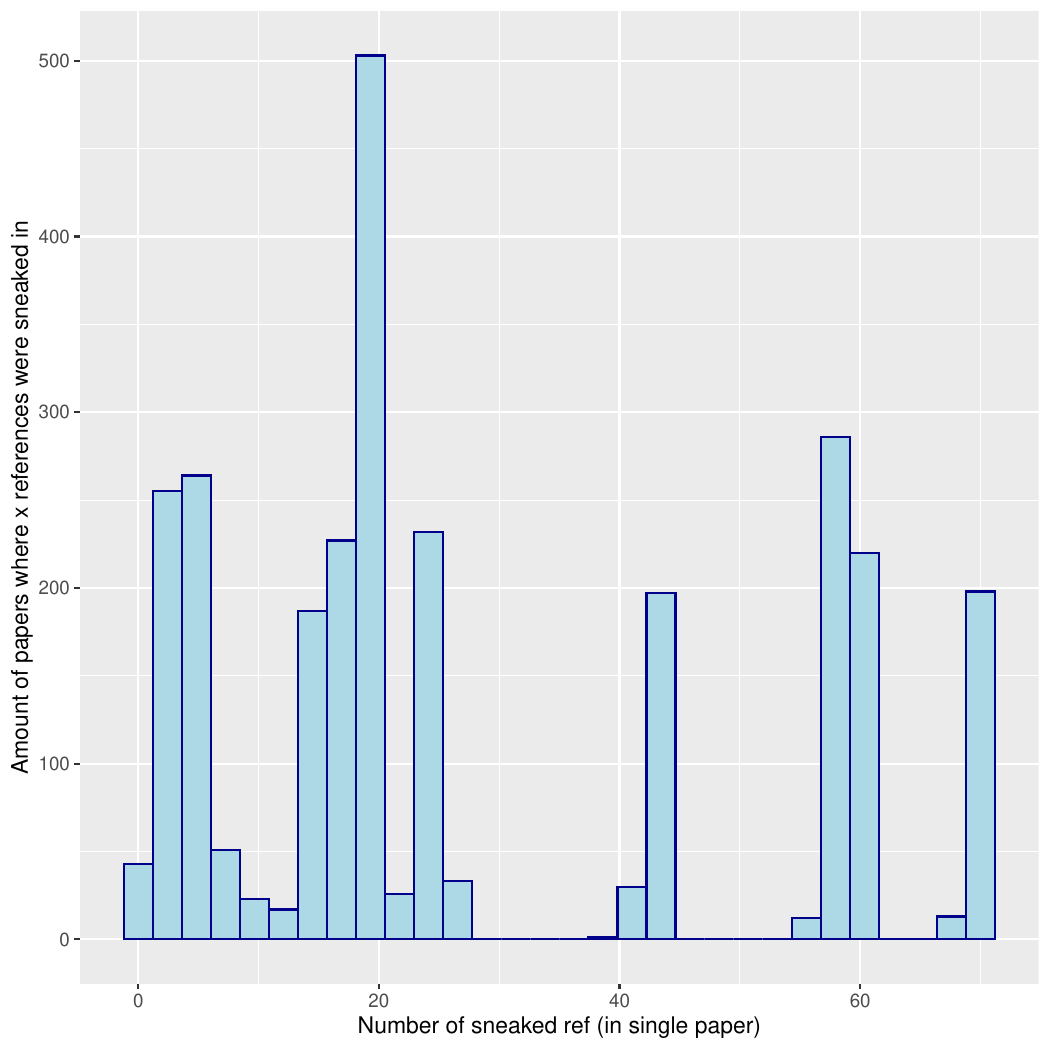}
		\caption{DOIs with at least one sneaked reference, $181$ DOIs have at least $70$ sneaked references.}
		\label{Fig:SneakedPerPaperBarZoom}
	\end{subfigure}
  \captionsetup{width=.8\linewidth}
	\caption{How many DOIs have $x$ sneaked references. 
	The mode (the most frequent value) is around $20$ sneaked references.
	}
	\label{Fig:SneakedPerPaperBarBoth}
\end{figure}

\begin{itemize}
    \item The corpus is composed of $4,077$ DOIs prefixed by \href{https://www.ijisrt.com}{10.38124/ijisrt}.

    \item For $3,222$ DOIs, a non-empty reference list $\mathcal R_{C}$ was downloaded from {\tt api.crossref}.  
	An empty result can reflect the fact that a document does not contain any reference list.
	But it also happens when the publisher did not register any reference list for this particular DOI.
	This means that the references listed in the real document are \emph{lost} \LO{} \citep[see][]{Besan2024Sneaked}. 
	Despite being present in the PDF, they are not registered with Crossref and are not credited to the cited document.
	Consequently, for $4,077-3,222=855$ DOIs the number of sneaked references is zero, as strictly no references are registered with Crossref.

    \item For $3,940$ DOIs, a non-empty reference list $\mathcal R_{G}$ has been extracted by Grobid from the PDF files.
	An empty list is generated either when the document does not contain any reference list or when Grobid failed to identify the reference section.
	
    \item The records for $3,132$ DOIs have both $\mathcal R_{C}\neq \emptyset$ and $\mathcal R_{G}\neq\emptyset$. They contain a total of $78,736$ sneaked references.

    \item Overall, the $80,205$ sneaked references were found in $2,782$ Crossref records. 
    The number of sneaked references in a single paper ranges from $1$ to $71$, with an average of $28.83$ sneaked references par paper (see distributions in \autoref{Fig:SneakedPerPaperBarBoth}). 

    \item The sneaked references are benefiting $2,703$ different DOIs. The most extreme case is a single DOI benefiting from $6,059$ undue citations.

    \item DOIs with sneaked references were created with Crossref between March 2024 and November 2024.
\end{itemize}

\subsection{Per beneficiary analysis}
\label{subsec:benef}

\begin{figure}[htbp]
\centering
	\begin{subfigure}{.45\textwidth}
	  	\centering
		\includegraphics[width=\linewidth]{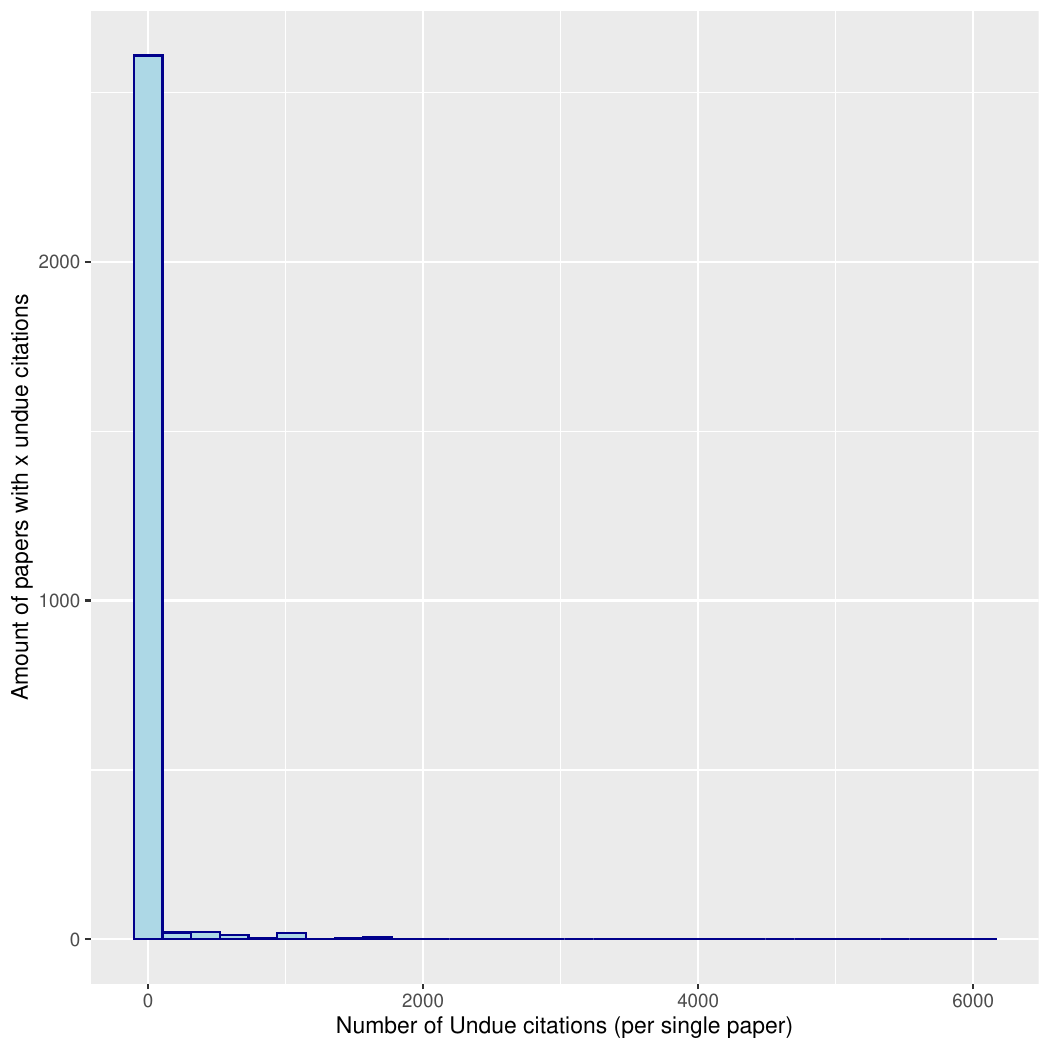}
		\caption{All DOIs, including those ($2,469$) that benefit from a single undue citation.}
		\label{Fig:BenfBarAll}
	\end{subfigure}
	\hspace{0.25cm}
	\begin{subfigure}{.45\textwidth}
	  	\centering
		\includegraphics[width=\linewidth]{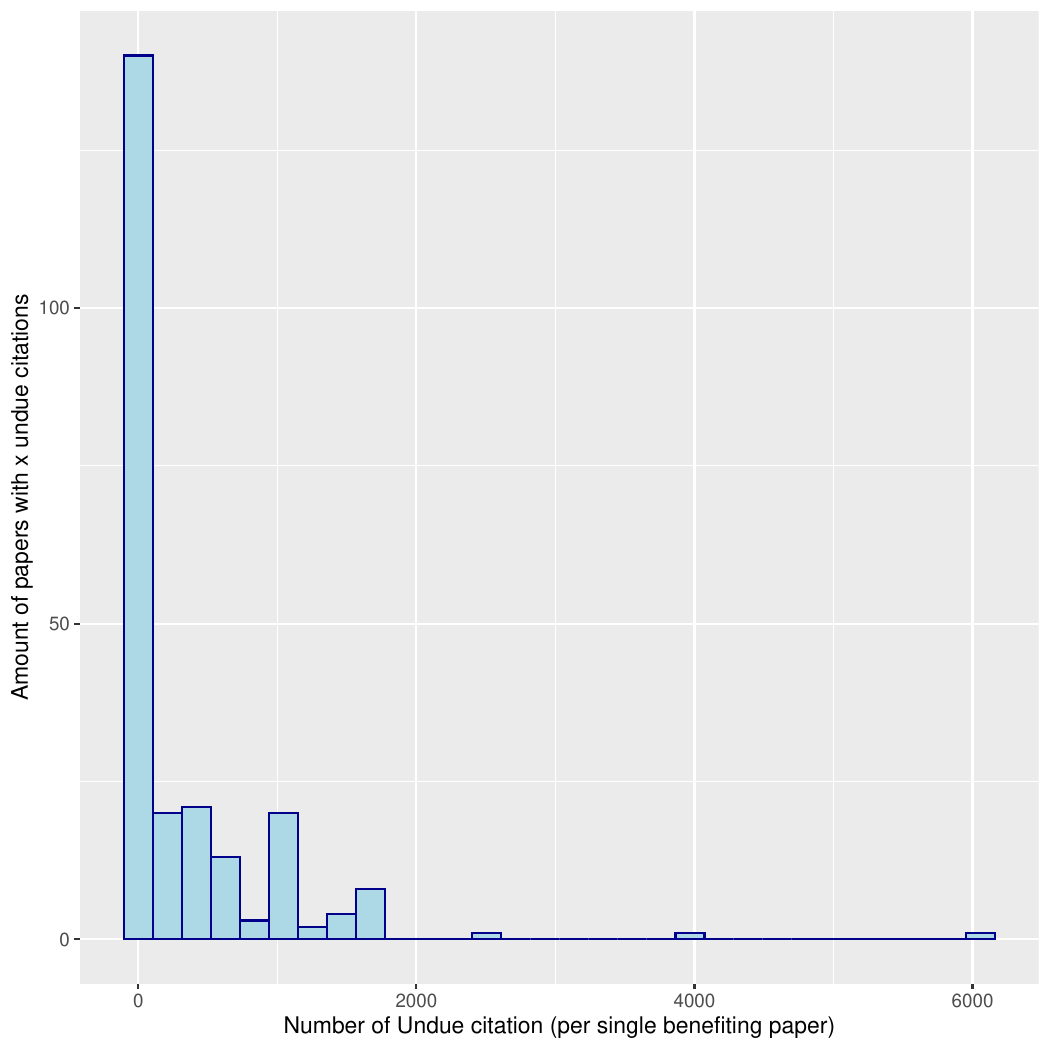}
		\caption{Only DOIs benefiting from more than one sneaked reference.}
		\label{Fig:BenfBarAllZoom}
	\end{subfigure}
  \captionsetup{width=.8\linewidth}
	\caption{How many DOIs benefit from $x$ undue citations? 
	For example, $2,607$ DOIs benefited from $1$ to $29$ undue citations
, while only $138$ DOIs are benefiting from $2$ to $31$ undue citations.
	}
	\label{Fig:BenfBar}
\end{figure}

\begin{figure}[htbp]
\centering
\includegraphics[width=0.7\textwidth]{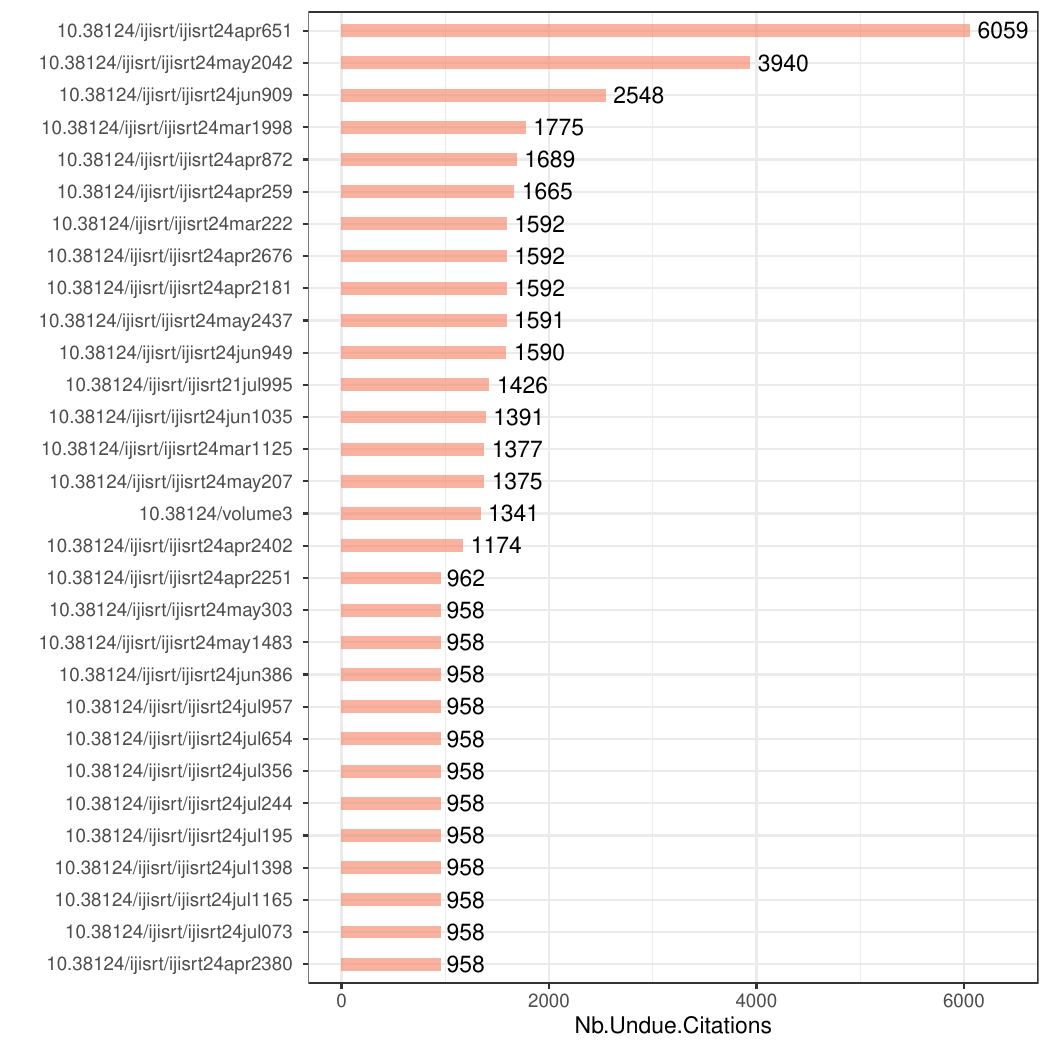}
\caption{\label{fig:BenfBarChart} The $30$ DOIs that benefit the most from sneaked references.}
\end{figure}

A total of $80,205$ sneaked references are benefiting $2,782$ different DOIs.
The average count of undue citations per benefiting DOI is $28.83$.
Nevertheless, the distribution is very \emph{unbalanced} as show in \autoref{Fig:BenfBar}. 
The overwhelming majority of DOIs ($n=2,469$) are credited with only a single undue citation. 
On the other hand, a small number of DOIs benefit from a significant number of undue citations.

The $30$ DOIs that benefit the most from sneaked references are shown in \autoref{fig:BenfBarChart}. 
The figure also shows the count of undue citations that these DOIs benefit from.
The DOI benefiting the most (\doi{10.38124/ijisrt/ijisrt24apr651}) from sneaked references is credited with $6,059$ undue citations.
Consequently, this particular DOI is incorrectly credited with $1.8k$ and 1.7k citations by OpenAlex and Dimensions, respectively (See \autoref{Fig:Dim} and \autoref{Fig:OpenAlex}).
This shows that some of the sneaked references effectively made their way through the counting processes onto some scientometric platforms.

\subsection{Time Analysis} 

The information available from Crossref includes the date on which a DOI was first registered: {\tt creation date}.
For sneaked references, we decided to compare the creation date of the citing DOI and the creation date of the cited DOI. 

\subsubsection{When were benefiting DOIs created?}

\begin{figure}[htbp]
\centering
	\begin{subfigure}{.45\textwidth}
	  	\centering
		\includegraphics[width=\textwidth]{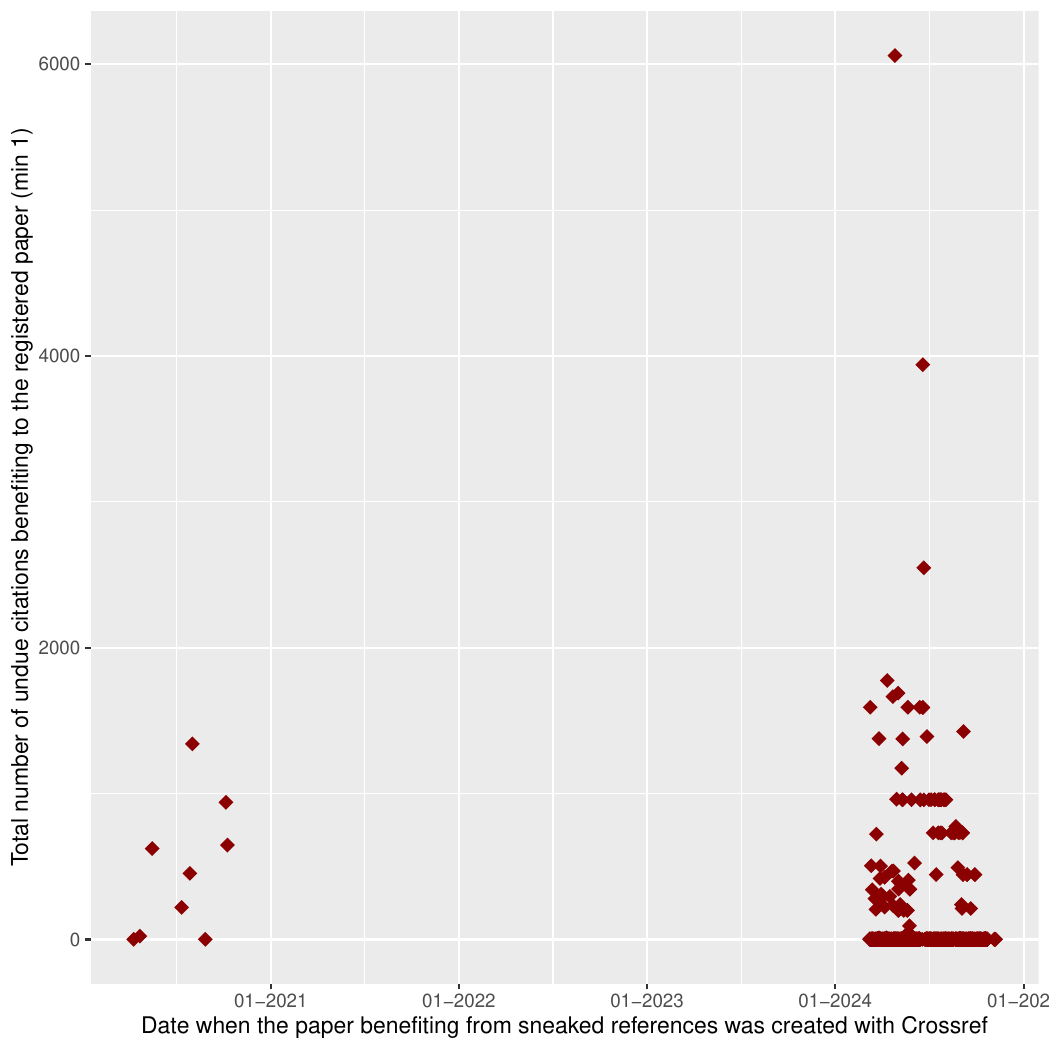}
		\caption{The observation on the extreme left represents sneaked references benefiting \doi{10.38124/volume4issue7}, a whole volume published in April 2020.}
		\label{UndueVSTime}  
	\end{subfigure}
	\hspace{0.25cm}
	\begin{subfigure}{.45\textwidth}
	  	\centering
		\includegraphics[width=\textwidth]{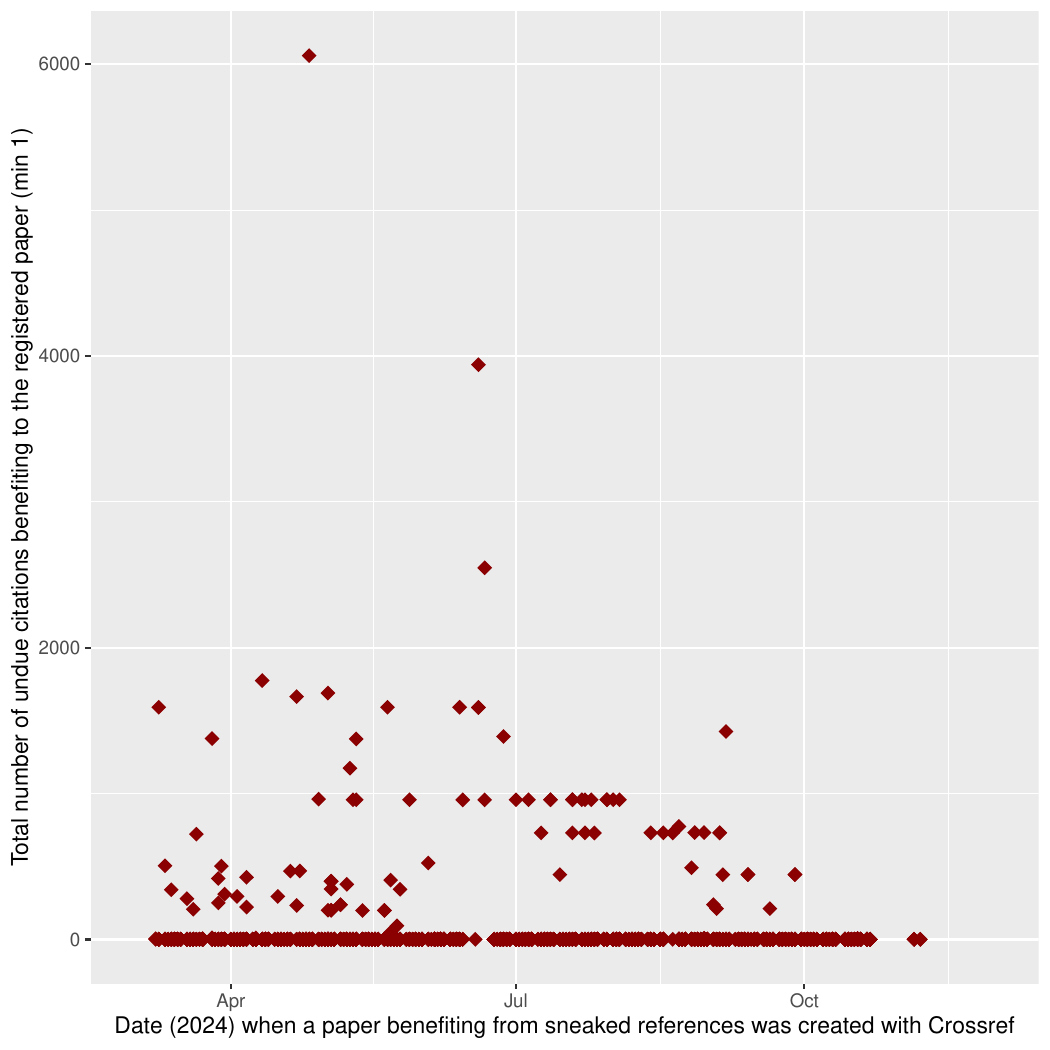}
		\caption{Same as the left panel but zoomed in on the year 2024. The DOI \doi{10.38124/ijisrt/ijisrt24apr651} benefiting the most from sneaked references ($6,059$) was created in April 2024.}
		\label{UndueVSTimeZoom}
	\end{subfigure}
	\caption{Number of sneaked references with regards to when the benefiting DOI was created with Crossref.}
\end{figure}

The oldest unduly cited DOI is \doi{10.38124/volume4issue7} which is the identifier of a volume published in April 2020 (See~\autoref{UndueVSTime}). 
It seems that this citation does not benefit any individual papers of the volume.

All but nine of the benefiting DOIs have been created between 2024-03-08 at 12:14 and 2024-11-07 at 12:54.
The DOI that benefits the most from sneaked references ($6,059$ undue citations) has been published in April 2024 (see \autoref{UndueVSTimeZoom}).

\subsubsection{When were DOIs with sneaked references created?}

\begin{figure}[htbp]
\begin{center}
\includegraphics[width=.5\textwidth]{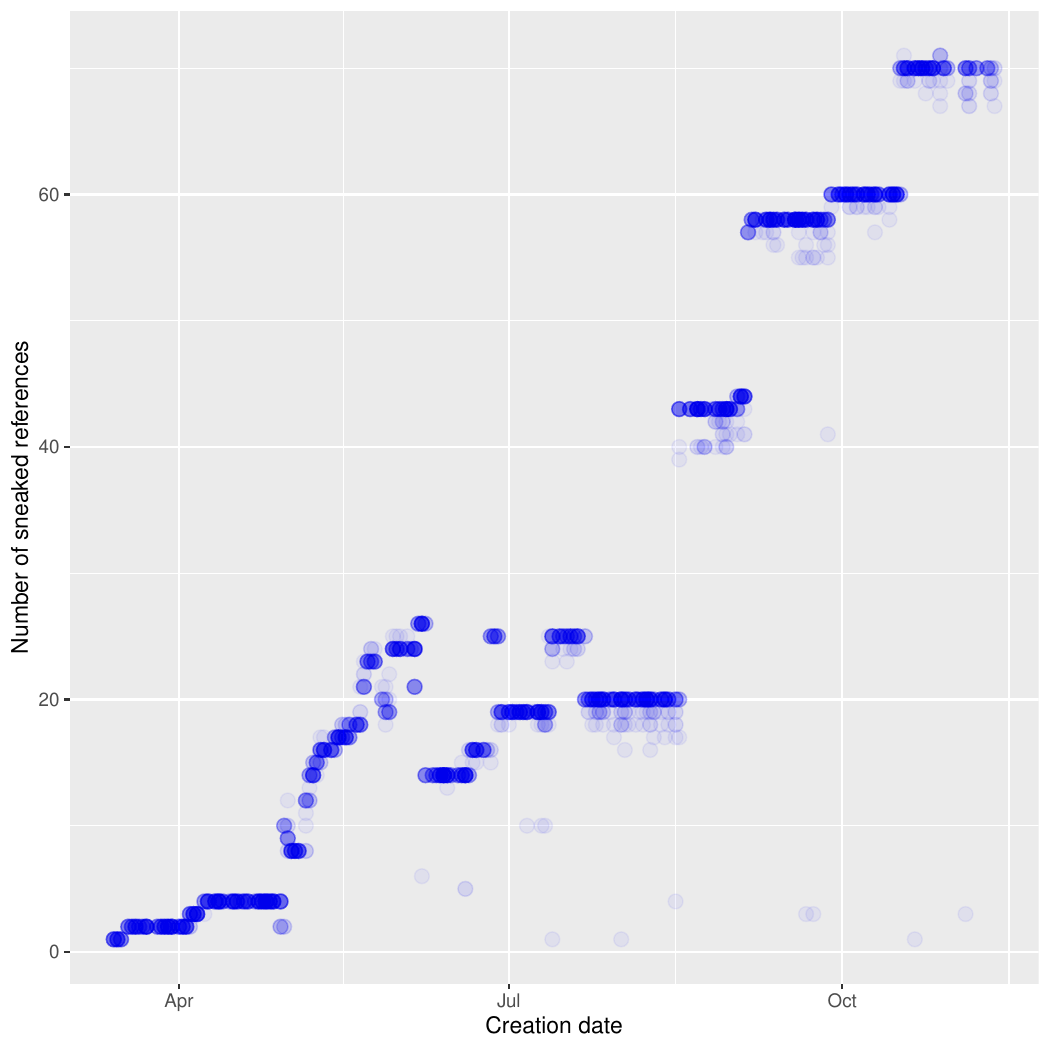}
 \captionsetup{width=.8\linewidth}
\caption{Date when DOIs with sneaked references were created with Crossref. The more intense the color, the more DOIs with the same number of sneaked references are registered at that time.}
\label{DateSneakedTrans}
\end{center}
\end{figure}

According to Crossref metadata, DOIs with sneaked references were created between the 2024-03-14 and the 2024-11-25 (see \autoref{DateSneakedTrans}).

The first 14 records featuring sneaked references have been registered with only one of them, on 2024-03-14. These individual sneaked references benefit different DOIs. 

The number of sneaked references per DOI increased quite rapidly to reach a maximum of 71.
The 6 DOIs with 71 sneaked references were all created on either 2024-10-28 or 2024-10-18.

\subsubsection{Coherence between citing and cited creation date}

Since sneaked references began to be included on 2024-03-14, and all but one of the beneficiaries started to be created 5 days earlier (2024-03-09), it is important to check the temporal coherence between the creation dates of the citing and cited works.

\begin{figure}[htbp]
\centering
	\begin{subfigure}{.45\textwidth}
	  	\centering
		\includegraphics[width=\textwidth]{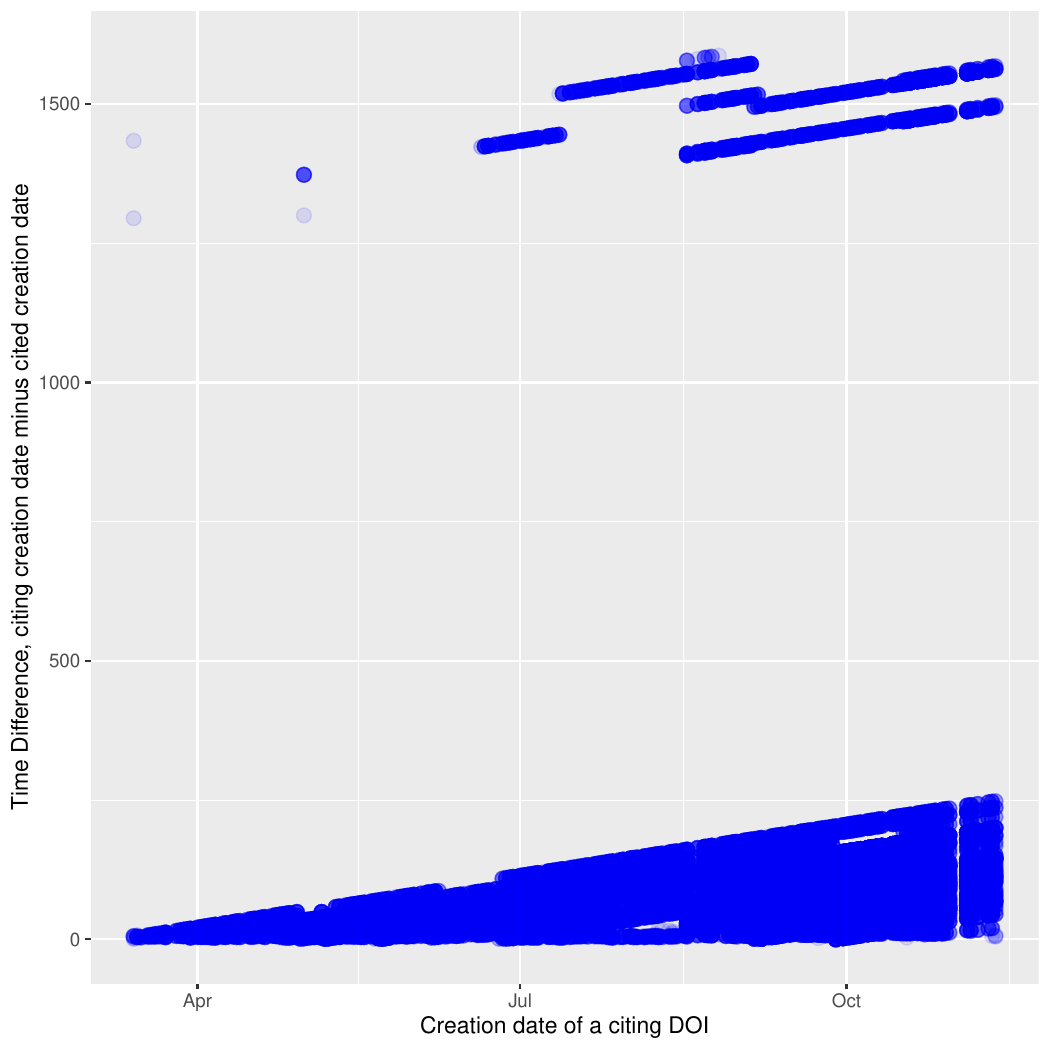}
		\caption{The light blue outliers in the upper left corner are the sneaked references to the oldest unduly cited DOIs, sneaked in on 2024-03. The upper band are sneaked references to `old' DOIs created in 2020 (see \autoref{UndueVSTime}).}
		\label{fig:Coherence}
	\end{subfigure}
	\begin{subfigure}{.45\textwidth}
	  	\centering
		\includegraphics[width=\textwidth]{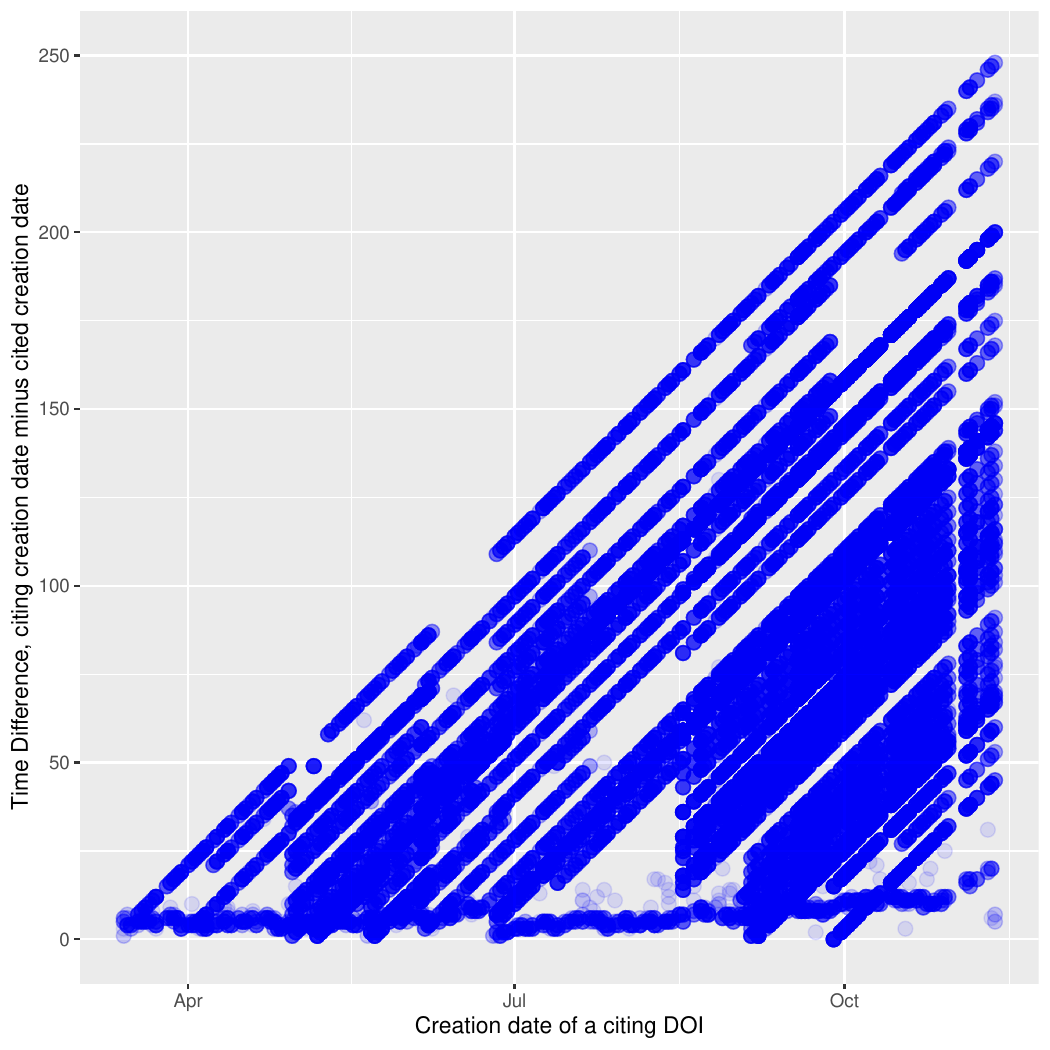}
		\caption{Same as panel (a), zoom with the upper band removed.\\~\\~\\}
		\label{fig:CoherenceZoom}
	\end{subfigure}
  \captionsetup{width=.8\linewidth}
	\caption{Temporal coherence between citing and cited DOIs. The $y$ axis is the number of days between the creation date of the cited DOIs and the creation date of the citing DOIs ($x$ axis). The darker the blue is, the more observations there are.}
	\label{fig:TempCO}
\end{figure}

\autoref{fig:TempCO} shows the number of days between the `citing' creation date and the `cited' creation date.
This number is always positive showing that, for all sneaked references, the citing DOIs were created after the cited ones.

Nevertheless, time differences are quite small, starting from $0$ days (one instance on 2024-05-22), and are slowly increasing as times passes by (see~\autoref{fig:CoherenceZoom}).

\begin{figure}[htbp]
\centering
	\begin{subfigure}{.45\textwidth}
	  	\centering
	  	\includegraphics[width=\textwidth]{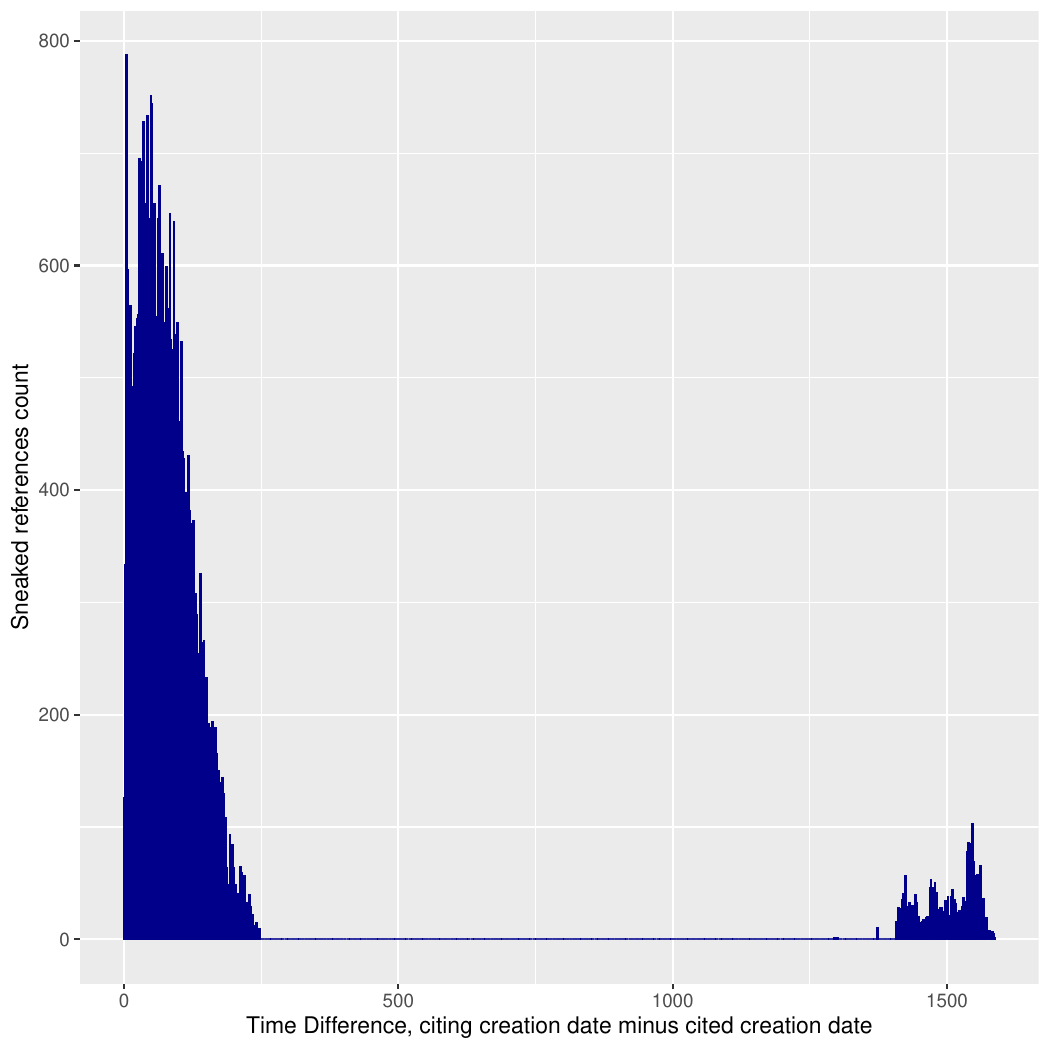}
        \captionsetup{width=.8\linewidth}
        \caption{Distribution of time differences between the cited and citing paper for sneaked references.}
        \label{fig:TimeDiffBar}
	\end{subfigure}
	\begin{subfigure}{.45\textwidth}
	  	\centering
	  	\includegraphics[width=\textwidth]{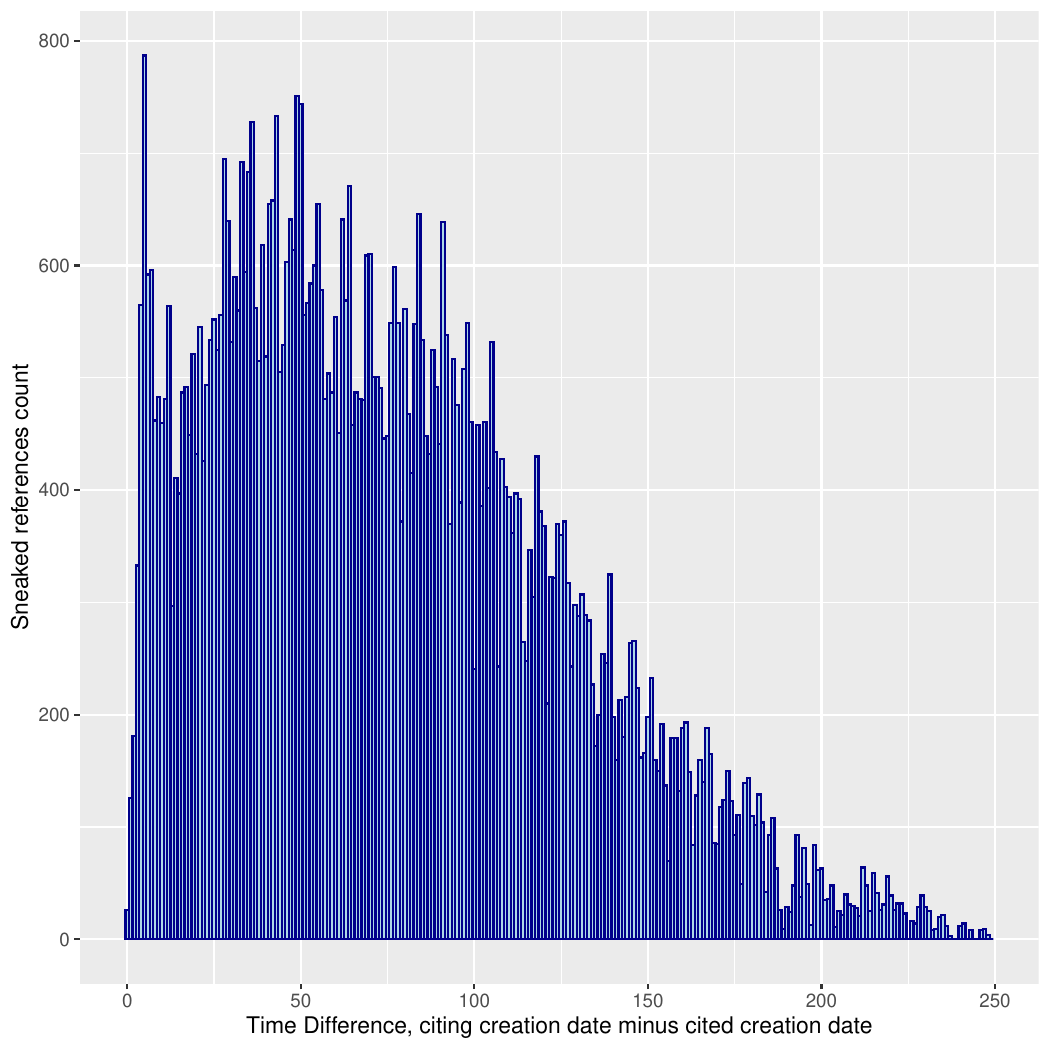}
        \captionsetup{width=.8\linewidth}
        \caption{Distribution of time differences (Zoom) between the cited and citing paper for sneaked references.}
        \label{fig:TimeDiffBarZoom}
    \end{subfigure}
    \caption{Distribution of time differences between the cited and citing paper for sneaked references.}
\end{figure}

\autoref{fig:TimeDiffBarZoom} shows the time difference distribution ($0\leq \delta \leq 250$ in days). The most frequent (with $\sim$800 occurrences) value is a difference of six days between citing and cited DOIs\ldots{} Half of the sneaked references are citing DOIs that were created less than 73 days (median) before their own creation.

\subsection{Summarizing evidence}

As a result it can be said that sneaked references were first added in small numbers in April 2024.
At that time only a handful of references were unduly added at registration time. 
The number of sneaked references increased little by little reaching a maximum of $71$ sneaked references per paper in November 2024.

The sneaked references are all benefiting the journal they are sneaked in and thus benefiting the publisher that registered them. 

The time difference between the cited and the citing paper, for sneaked references, is often surprisingly small.
It is also worth noting that sneaked references are not benefiting to cited papers in a very unbalanced way. 
Most of the papers are only benefiting from one single sneaked reference, while a few are benefiting for hundreds of undue citation.

In the light of these data, it is hard to find definitive evidence that differentiates intentional manipulation from genuine malfunctions in the meta-data registration process.

Nevertheless, awkward metadata registered with Crossref might help to identify venues where references are sneaked in.
The next section investigates this hypothesis.

\section{Attempts to detect sneaked references at a large scale}
\label{sec:Large}

\autoref{heuristic} explains how sneaked references can be discovered when they occur together with duplicated references.
\autoref{MatScale} discuses results of an attempt to identify sneaked references at a large scale by applying method $\mathcal M_0$ on 47,170,721 documents.

\subsection{Duplicate based Heuristic to circumscribe sneaked references}
\label{heuristic}

We hypotheses that sneaked references appear together with duplicated references.
Therefore, we try to detect groups of duplicated references in the hope to circumscribe sneaked references.

A bibliography should not contain the same \texttt{DOI} multiple times.
The approach we adopted is to consider only the \texttt{DOI} of references to identify duplicates.  
Thus, two references in a same citing article pointing to the same \texttt{DOI} is considered as a \emph{duplicate reference}.
Working at the \texttt{DOI} level to spot duplicates is imperfect. 
We noticed that some DOIs occur multiple times in a Crossref record of a bibliography that does not feature duplicates. 
This can happen when a DOI of a book is used to identify different chapters of this book.
For this reason, we excluded books and book chapters from our analysis.

We used the latest Crossref snapshot downloaded on 23 November 2023. 
It contains a total of 991,206,078 reference entries including duplicates.
3,755,847 ($0.38\%$) of these are duplicated 1+ times.
Thus, the dump contains a total of 986,772,474 distinct references.
Overall, 4,433,404 ($0.45\%$) reference entries are duplicates; and there is an average of 1.18 duplicates per duplicated reference.

Our goal is to identify \emph{entities} that benefit the most from duplicated references. 
The entities we consider are either publishers, authors, articles, or journals.
Let us introduce the following notations:
\begin{itemize}
    \item 
    Let $D$ be the set of documents (limited to articles),
    \item 
    $A$ be the set of authors,
    \item 
    $P$ be the set of publishers
    \item 
    and $J$ be the set of journals.
\end{itemize}

Authors are authoring documents that are published in journals and documents are referencing other documents. To denote this, we'll use the following notations: 
\begin{itemize}
    \item For $a \in A$ and $d \in D$, “$\opauth{a}{d}$” means: $a$ is author of $d$
    \item For $j \in J$ and $d \in D$, “$\opjrn{d}{j}$” means: $d$ is published in $j$
    \item For $(d_1,d_2) \in D^2, n \in \mathbb{N}$, \\
    “$\opciten{d_1}{d_2}{n}$” means: $d_1$'s metadata contains exactly $n$ references to $d_2$. \\
    “$\opcite{d_1}{d_2}$” means: $d_1$'s metadata contains at least one reference to $d_2$.
\end{itemize}
We also consider the following sets:
\begin{itemize}
    \item $D_j$ the set of the documents published in journal $j \in J$: \(D_j = \left\{d \in D \mid \opjrn{d}{j}\right\}\).
    \item $R_d$ the set of the references of document $d \in D$: \(R_d = \left\{\left(d,r\right) \in D^2 \mid \opcite{d}{r}\right\}\).
    \item $R^+_d$ the set of references of document $d \in D$, that are duplicated 1+ times:
    
\(R^+_d = \left\{\left(d,r\right) \in R_d \mid n \in \mathbb{N}, n > 1, \opciten{d}{r}{n}\right\}\). Note that $R^+_d \subseteq R_d$.
    \item $C_d$ the set of the documents citing document $d \in D$:\(C_d = \left\{c \in D \mid \opcite{c}{d}\right\}\). 
    
\end{itemize}

To measure how much a registered DOIs contains duplicated references or how much a paper benefit from duplicated references, the following measure are defined:
\begin{itemize}
    \item $\nbref{d_1}{d_2}$ denotes the \textbf{number of times} $d_1 \in D$ contains a reference to $d_2 \in D$.
    \item $ Benef^+(d) = \sum_{c \in C_d}\left(\nbref{c}{d}-1\right)$ is the number of duplicated references benefiting $d$
    and \\
    $ Benef(d) = \left|\left\{c \in C_d \mid \nbref{c}{d}>1\right\}\right|$ is the number of 1+ duplicated references $d$.
    \item $ NbRefDup^+(d) = \sum_{\left(d,r\right) \in R^+_d}\left(\nbref{d}{r}-1\right)$ is the number of duplicated  references in metadata registered for document $d$,
    \\
    $ NbRefDup(d) = \left|R^+_d\right|$ is the number of reference duplicated 1+ times, in metadata registered for document $d$.
\end{itemize}

\begin{table}[ht]
    \centering
        \centering
        \begin{tabular}{l
                        S[table-format=2.0]
                        S[table-format=2.0]}
            \toprule
            \multicolumn{1}{c}{\texttt{DOI} of the cited document $d$} & \multicolumn{1}{c}{$Benef^+(d)$} & \multicolumn{1}{c}{$Benef(d)$} \\\midrule
            \mydoi{10.17265/2159-5313/2016.09.003} & 10994 &  6147\\
            \mydoi{10.1109/geoinformatics.2015.7378602} & 2042 & 464\\
            \mydoi{10.1038/scientificamerican0703-56}   & 696  & 1\\
            \mydoi{10.1089/glre.2016.201011}            & 657  & 336\\
            \mydoi{10.4064/fm-146-3-215-238}            & 504  & 229\\\bottomrule
        \end{tabular}
        \caption{$Benef^+(d)$ the number of duplicated references benefiting article $d$.
        $Benef(d)$ the number of 1+ duplicated references to each document.
        }
        \label{tab:most-popular-articles-dupli}
    \end{table}

    \begin{table}[ht]
        \centering
        \begin{tabular}{l
                        S[table-format=2.0]
                        S[table-format=2.0]}
            \toprule
            \multicolumn{1}{c}{\texttt{DOI} of the citing document $d$} & \multicolumn{1}{c}{$NbRefDup^+(d)$} & \multicolumn{1}{c}{$NbRefDup(d)$}\\
            \midrule
            \mydoi{10.1190/segam2016-full} & \bf 1029 & 485\\
            \mydoi{10.2903/sp.efsa.2017.en-1246} & 1020 & \bf 815\\
            \mydoi{10.1190/segam2016-full2}  & 919  & 470\\
            \mydoi{10.14412/1995-4484-2020-191-197} & 863  & 62\\
            \mydoi{10.4236/abb.2012.324065} & 696  & 1\\
            \bottomrule
        \end{tabular}
        \caption{
        Top five documents for
        $NbRefDup^+(d) $ the number of duplicated  references in metadata registered for document $d$.
    $ NbRefDup(d)$ is number of reference duplicated 1+ times, in metadata registered for document $d$.
        }
        \label{tab:most-duplications-articles}
\end{table}

Some articles ‘benefit’ from an impressive number of duplications.
The top 5 is shown in \autoref{tab:most-popular-articles-dupli}). 
Quite interestingly, the landing page (using doi.org) for the first row is currently a generic error page. 
\autoref{tab:most-duplications-articles} provide the top 5 papers for which metadata contains a lot of duplicated references.
Again the landing page for the first DOI leads to a \emph{Page not Found} error.
Visual inspection of Crossref records for these five DOI reveal simple duplicated references without any obvious sneaked references.

To measure how much a journal does contain duplicated references, the following measures are computed:
\begin{itemize}
    \item $JourDup^+(j)= \sum_{d \in D_j}\sum_{\left(d,r\right) \in R^+_d}\left(\nbref{d}{r}-1\right)$ the number of duplicated references found in the metadata of a journal $j$.
    \item $ JourDup(j) = \left|\left\{d \in D_j | R^+_d \neq \emptyset\right\}\right|$ the number of documents registered for journal $j$ that contain at least one duplicated reference.
\end{itemize}

\begin{table}[ht]
    \centering
    \resizebox{0.95\textwidth}{!}{ 
    \begin{tabular}{p{10cm}S[table-format=7.0]S[table-format=7.0]S[table-format=3.0]}
        \toprule
        \multicolumn{1}{c}{Journal $j$} & \multicolumn{1}{c}{$JourDup^+(j)$} & \multicolumn{1}{c}{$JourDup(j)$} & \multicolumn{1}{c}{$JourDup^+(j)/JourDup(j)$}\\
        \midrule
        \emph{SSRN Electronic Journal} & \bf 110390 & \bf 40082 & 2.8\\
        \emph{Journal of Behavioral Addictions} & 22286 & 603 & 37.0\\
        \emph{RSC Advances} & 21378 & 13316 & 1.6\\
        \emph{The Journal of Contemporary Dental Practice} & 20407 & 1731 & 11.8\\
        \emph{Internationa Journal of Sports Physiology and Performance} & 19351 & 1020 & \bf 19.0\\
        \emph{Health} & 18966 & 1412 & 13.4\\
        \emph{Scientific Reports} & 17856 & 14456 & 1.2\\
        \emph{PLOS ONE} & 17541 & 13216 & 1.3\\
        \emph{American Journal of Plant Sciences} & 16453 & 1226 & 13.4\\
        \emph{Creative Education} & 16355 & 863 & \bf 19.0\\
        \bottomrule
    \end{tabular}
    
    }
    \caption{ The ten journals that registered the most duplicated references with Crossref. $JourDup^+(j)$ is the total number of duplicated references registered by this journal, $JourDup(j)$ the number of documents registered for journal $j$ containing at least one duplicated reference. 
    }
    \label{tab:publishers-per-duplications}
\end{table}

\autoref{tab:publishers-per-duplications} lists journals for which the highest number of duplicated references are found.
Such journals push a great amount of duplicated reference metadata to Crossref.
It is important to note that, despite the name, \emph{SSRN Electronic Journal} is a platform for pre-prints. 
The first position might be explained by the high number of papers the platform register with Crossref.
In second position, the \emph{Journal of Behavioral Addictions} did register an average of 37 duplicated references over $603$ articles.
Again, close inspection of some examples reveals only simple duplicated references without any obvious sneaked references.

Previous work~\citep{Besan2024Sneaked} showed that sneaked references are sometimes benefiting to particular authors.
Thus, identifying authors that benefit from duplicated references may also be the ones that benefit from sneaked references.

We note $s_{1a}(j,a)$ the proportion of references found in journal $j$ that cite a paper authored by $a$:
$$s_{1a}(j,a) = \frac{\sum_{d \in D_j}\left|\left\{\left(d,r\right) \in R_d \mid \opauth{a}{r}\right\}\right|}{\sum_{d \in D_j}\left|R_d\right|}$$
$s_{1b}(j,a)$ refers to the proportion of duplicated references found in journal $j$ that cite a paper authored by $a$:
$$s_{1b}(j,a) = \frac{\sum_{d \in D_j}\left|\left\{\left(d,r\right)  \in R^+_d \mid \opauth{a}{r}\right\}\right|}{\sum_{d \in D_j}\left|R^+_d\right|}$$
Hypothesising duplications occur randomly because they are mistakes, we should observe $s_{1a}(j,a) \sim s_{1b}(j,a)$. If $s_{1b}(j,a)$ is far greater than $s_{1a}(j,a)$ this mean that duplicated references in journal $j$ are benefiting in an abnormal proportion to author $a$.

That is why we compute $s_1(j,a)$ an estimation of the number of duplicated references in journal $j$ that are \emph{statistically speaking} unexpected for authors $a$:
$$ s_1(j,a) = (s_{1a}(j,a) - s_{1b}(j,a)) \cdot \sum_{d \in D_j}\left|\left\{\left(d,r\right) \in R^+_d \mid \opauth{a}{d}\right\}\right| $$

\begin{table}[ht]
    \centering
    \resizebox{1\textwidth}{!}{ 
    \begin{tabular}{p{7.5cm}p{3.8cm}S[table-format=3.0]S[table-format=3.0]S[table-format=3.0]S[table-format=3.0]S[table-format=3.0]}
        \toprule
        \multicolumn{1}{c}{Journal} & \multicolumn{1}{c}{Author} & \multicolumn{1}{c}{\makecell{No. dupli.\\in journal\\to author}} & \multicolumn{1}{c}{\makecell{No. dupli.\\in journal}} & \multicolumn{1}{c}{\makecell{No. ref.\\from journal\\to author}} & \multicolumn{1}{c}{\makecell{No. ref.\\in journal}} & \multicolumn{1}{c}{Score}\\
        \midrule
        \emph{Econometrics: Alchemy or Science?} & david f hendry & 76 & 104 & 204 & 1391 & 44.4\\[5pt]
        \emph{Construction and Architecture} & timofey krakhmalnyy & 35 & 49 & 58 & 1273 & 23.4\\[5pt]
        \emph{International Journal of Scientific Research in Science and Technology} & harikriishna b jethva & 142 & 981 & 213 & 18990 & 19.0\\[20pt]
        \emph{International Journal of Scientific Research in Science and Technology} & bhavesh kataria & 142 & 981 & 242 & 18990 & 18.7\\[20pt]
        \emph{International Journal of Laser Dentistry} & a l mckenzie & 75 & 315 & 75 & 7653 & 17.1\\[5pt]
        \emph{Construction and Architecture} & sergej evtushenko & 25 & 49 & 57 & 1273 & 11.6\\[5pt]
        \emph{International Journal on Disability and Human Development} & daniel t l shek & 199 & 2650 & 283 & 10303 & 9.5\\[20pt]
        \emph{International Journal on Applied Engeneering and Management Letters} & p s aithal & 54 & 182 & 551 & 4154 & 8.9\\[20pt]
        \emph{Berichte der deutschen chemischen Gesellschaft} & h staudinger & 213 & 3901 & 1027 & 56035 & 7.7\\[20pt]
        \emph{An Introduction to Community and Primary Health Care} & elizabeth halcomb & 36 & 164 & 64 & 2224 & 6.9\\[20pt]
        \emph{Cambridge Handbook of Multimedia Learning} & richard e mayer & 49 & 216 & 227 & 2455 & 6.6\\[5pt]
        \emph{Bears of the World} & jon e swenson & 114 & 1071 & 249 & 4602 & 6.0\\[5pt]
        \bottomrule
    \end{tabular}
    
    }
    \caption{Pairs of authors and journals sorted by $s_1(j,h)$ score. The columns reflect: name of journal, name of author, number of references that are duplicated 1+ times in journal and benefiting to author, number of 1+ duplicated reference(s) in journal, number of references (without duplications) benefiting to author in journal, total number of references (without duplications) in journal.}
    \label{tab:req-sql-jules}
\end{table}

Computing this score (\autoref{tab:req-sql-jules}) can reveal statistical anomalies that may (or may not) reflect citation gaming. 

The computed leaderboard features Harikrishna~B.~\textsc{Jethva} and Bhavesh \textsc{Kataria} together with the \emph{International Journal of Scientific Research in Science and Technology} from the \emph{Technoscience Academy} publisher. 
This case being the one described in~\citep{Besan2024Sneaked}.
This result is coherent with our hypothesis that duplicated references might be correlated with sneaked references.
At the time of the Crossref snapshot was created metadata were not yet corrected.
Since then, when asked by Crossref, the publisher did correct the records and removed sneaked references.

Some other authors in this list are suspected to manipulate their $h$~index (e.g., P.~S.~\textsc{Aithal}\footnote{\url{https://www.researchgate.net/post/Excessive_self-citation_in_his_research_papers_which_has_artificially_inflated_his_H-index_score}}). 

We did not check all the articles published by the journal--author pairs of this leaderboard, and further analysis might give new interesting results.
Nevertheless, the case of the pair \emph{International Journal of Laser Dentistry} and A.~L.~\textsc{McKenzie} is of some interest.
we indeed found \emph{sneaked references} benefiting to A.~L.~\textsc{McKenzie}'s articles.
For example, \mydoi{10.5005/jp-journals-10022-1031} contains a duplicated \emph{sneaked reference} to \mydoi{10.1109/geoinformatics.2015.7378602}).
Deeper investigations reveal that these sneaked references are not resulting from intentional manipulations but most probably are the consequence of genuine errors.
This sneaked reference might be unintentional as its seems that this journal always sent the same reference list for all the metadata of its articles, except for the $n$ first references of each list that are replaced by the $n$ references of the current published article.
This reference list contains the expected list of references (found in the PDF file) but are always padded up to $155$ with the same set of sneaked references.
This journal is no more active and for each and every published article $d$ the website (landing page) is providing a list of exactly $155$ references that are the ones registered at Crossref.

\subsection{Scaling the detection of sneaked references to the entire scientific literature}
\label{MatScale}

Hoping that a combination of methods might enable the detection and validation of sneaked references at scale, we compared the length of the reference list extracted from PDFs using Grobid ($\mathcal M_0 $) with the references registered with Crossref.
For articles published since 2000, the number of reference items identified by Grobid were compared to the number of references provided by Crossref. In order to account for a reported Grobid uncertainty rate of 0.05, we allowed for a margin of error, comparing the length of references identified by the full-text Grobid processing to 0.95 times the Crossref references count.
A total of 4,172,499 articles out of 47,170,721 processed were found to have fewer references than 95 percent of the Crossref reference count.

In order to determine which publications had been added and which authors or journals had been inserted erroneously, we attempted to match the references extracted by Grobid with the identifiers supplied by Crossref.
This approach turns out to be challenging, partially as a result of the inconsistency of reference formatting in the PDFs (many were missing DOIs) but also highlighted an error in the initial approach: references which are provided in supplementary attachments were not counted in the original Grobid-processed PDFs.

Although this attempt to systematically identify erroneous references was not effective, there remain a total of 1,564,408 publications with between 5 and 500 additional references reported to Crossref beyond those identified by Grobid processing.
The possibility remains that this approach may work for a subset of the total dataset (see \autoref{perf:M0} for more details on the limitations of the precision of the extraction of references using Grobid).

\section{Conclusions}
\label{sec:conclu}

We investigated three ways ($\mathcal M_0$, $\mathcal M_1$, and $\mathcal M_2$) to automatically identify sneaked references by comparing references registered with Crossref and the ones extracted from PDF files using the Grobid software.

The first one $\mathcal M_0$ \citep{Besan2024Sneaked}, based the direct comparison between list lengths, leads to an overestimation (7\%) of the number of sneaked references.

The second one $\mathcal M_1$ relies on the last items of each list (\autoref{sec:meth1}).
It supposes than the order of the two lists is the same, which is quite a strong assumption.
If $Last_{C}=Last_{G}$ we conclude that $\mathcal R_{C}$ is correct with regards to the PDF version. 
This is always the case in this specific dataset.
But it could be that some references were still sneaked in, although that cannot be checked without a thorough manual analysis.

The third one $\mathcal M_2$ (\autoref{sec:meth2}) seems to provide the more accurate results.
It does not require any complex reference extraction from the PDF file,  

The current state of full text data prevents $\mathcal M_0$ from working at large scale. \\
Computing expensive so tried to limit the search using heuristic (duplication).

\subsection{Corrections and updates to references}
We identified a new set of sneaked references that benefit a single journal: \emph{IJISRT}.
The sneaked references are registered with Crossref along with the metadata for this journal.
As a result, sneaked references inflate citations counts for this journal and for some of its articles. 

Crossref provides infrastructure for registering metadata for scholarly works, including a DOI.
To use this infrastructure, organisations join Crossref as members, taking on related obligations\footnote{\url{https://www.crossref.org/membership/terms/}}. At the most basic level, Crossref members are responsible for depositing accurate metadata for each content item they produce.

When a serious issue with the metadata is detected, Crossref contacts the member to investigate the situation and work with them to rectify the problem where applicable. In some rare cases, the member’s access to register new scholarly works or update their existing records might be temporarily suspended or their membership permanently revoked.

The Crossref records for $\sim{}2.7k$ DOIs from the \emph{International Journal of Innovative Science and Research Technology} require corrections to remove the $\sim{}81k$ sneaked references. In November 2024, Crossref contacted the member responsible for the International Journal of Innovative Science and Research Technology to ask for an explanation. Based on the member's replies to the enquiries, it was clear there was an intention to manipulate the citation record, and as such Crossref have started the process to revoke this organization's  membership.

More information on Crossref’s membership revocation process can be found here \footnote{\url{https://www.crossref.org/operations-and-sustainability/membership-operations/revocation/}}. The most up-to-date list of revoked Crossref members is available here.\footnote{\url{https://docs.google.com/spreadsheets/d/1cCkdvtqEM1urmrUQZ4-LGz_Omf5812aVkRFJc5UryHw/edit}} This member will appear on the list if their revocation is ratified by the Crossref board.

Crossref encourages the community to report cases via the dedicated "metadata quality improvements" channel\footnote{\url{https://community.crossref.org/c/tech-support/metadata-quality-improve/45}} on its forum.

\subsection{Implications}

There are a number of possible sources of erroneous references, not all of which are nefarious. The identification of publications whose reference count fails to match the references actually listed in the references section (which may differ, in turn, from the in-text citations), is only the first, judgment neutral, step. There are some patterns detectable in the erroneous references, which hint at the source.

For example, in one situation there were several publications with a valid list of references which seemed to have been written on top of a constant, longer list of references in identical order. The number of total references was constant, while the number of valid references varied. This situation suggests to us an error in pasting, where a shorter list of references pasted over, rather than replaced, a longer list of references from an earlier metadata submission. In other cases, we found multiple identical references pasted at the end of the list of valid references. This behavior suggests a less technical source, and less benign intentions.

There are different methods for members to register scholarly metadata with Crossref.   
These range from plugins integrated into publishing platforms to registration forms with different metadata fields for members to fill. XML files can also by directly sent using HTTPS POST.
On the member side, during the publishing process different actors might have different kind of access to the metadata records, providing different kinds of opportunities to sneak references in or register erroneous metadata.

Sneaked references remain one of many possibilities to practice citation gaming (see \autoref{sec:intro}). 
More of such gaming will continue to prevail as long as academic value is tightly coupled to specific metrics. 

Given the variety of reasons for erroneous references, there are multiple approaches that could be taken to improve the situation, which include improving the tools by which editors submit reference lists, the automated deduplication of reference lists after submission, and systematically cross-checking publications using Grobid in collections (proprietary or otherwise) which contain full-text PDFs. Deliberate efforts, which measure the rate of success of these approaches, are advisable.

\subsection{Future work}
Beyond this specific data set, the extent to which sneaked references are distorting citation counts is unknown.
It might remain a very limited phenomenon but this needs to be verified by further investigations.

Identifying those journals or authors which have been most frequently associated with erroneous references, at scale, may allow us to identify the beneficiaries of sneaked references. This could act as a heuristic device to search for additional sneaked references (see \autoref{heuristic}), and to distinguish between erroneous and sneaked references.

Future work will attempt to identify erroneous references at a larger scale. We will continue to attempt to define patterns that can be used to flag sneaked references. This big data approach will help us determine whether the sneaked reference would have had a bibliometric effect, resulting in any increase in the Journal Impact Factor (or other journal-based metrics).

\paragraph{Data and code availability}

Along with the source code implementing $\mathcal M_1$ and $\mathcal M_2$ we are releasing the dataset, which can be found at \doi{10.5281/zenodo.14319568}.  
The code can be found at \doi{10.5281/zenodo.14291988}.

\paragraph{Conflicts of interest}

Two of the authors are employed by the providers of the data used in the analysis: Dimensions (KWB) and Crossref (DT).
GC is an AE at JASIST.

\paragraph{Acknowledgement}
CL and GC acknowledge the NanoBubbles project that has received Synergy grant funding from
the European Research Council (ERC), within the European Union’s Horizon 2020 program, grant agreement
no. 951393, doi:\doi{10.3030/951393}. LB was supported, in part by the Knut and Alice Wallenberg Foundation (grant KAW 2019\discretionary{.}{}{.}0024).
KWB would like to thank Balbir Thomas and Ruth Whittam for their participation in discovery and coding.

\bibliographystyle{apacite}
\interlinepenalty=10000 
\bibliography{references}

\section{Appendix}

\begin{figure}[htbp]
\centering
	\begin{subfigure}{.30\textwidth}
	  	\centering
		\includegraphics[width=\textwidth]{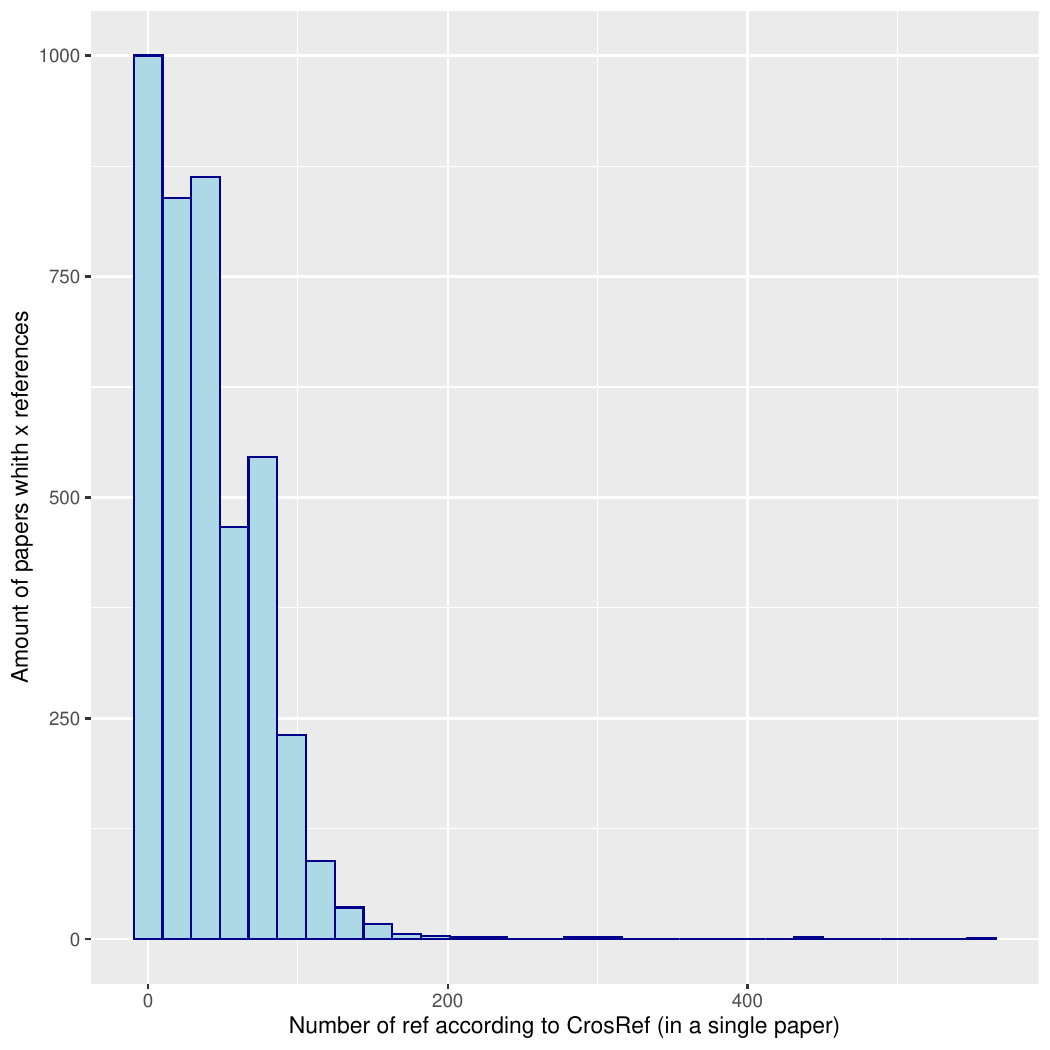}
		\subcaption{Distribution of the number of references registered for a DOIs with Crossref (length of $\protect\mathcal R_{C}$).}
		\label{fig:CrossRefBar}
	\end{subfigure}
	\hspace{0.25cm}
	\begin{subfigure}{.30\textwidth}
	  	\centering
		\includegraphics[width=\textwidth]{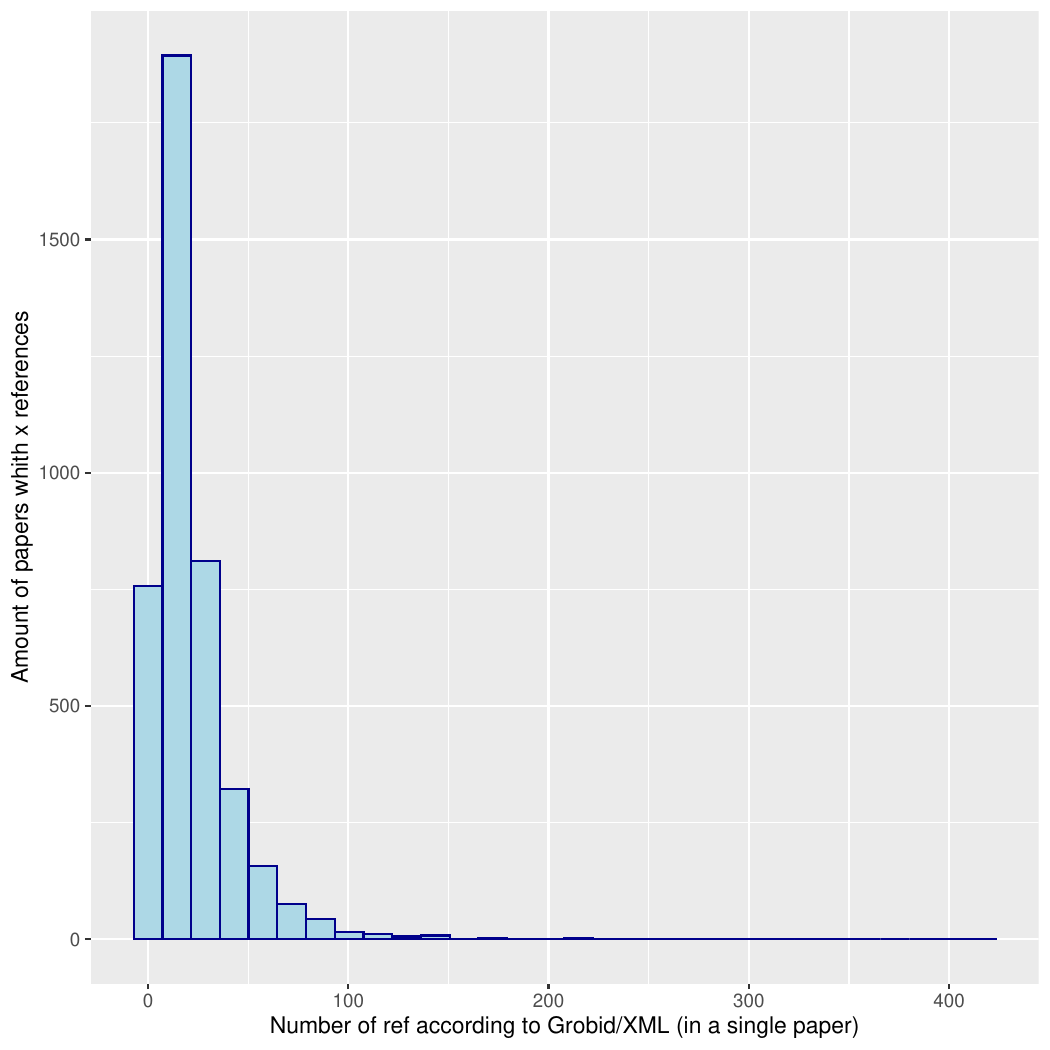}
		\subcaption{Distribution of the number of references extracted from the PDF (length of $\protect\mathcal R_{G}$).}
		\label{fig:DiffBar}
	\end{subfigure}
	\begin{subfigure}{.30\textwidth}
	  	\centering
		\includegraphics[width=\textwidth]{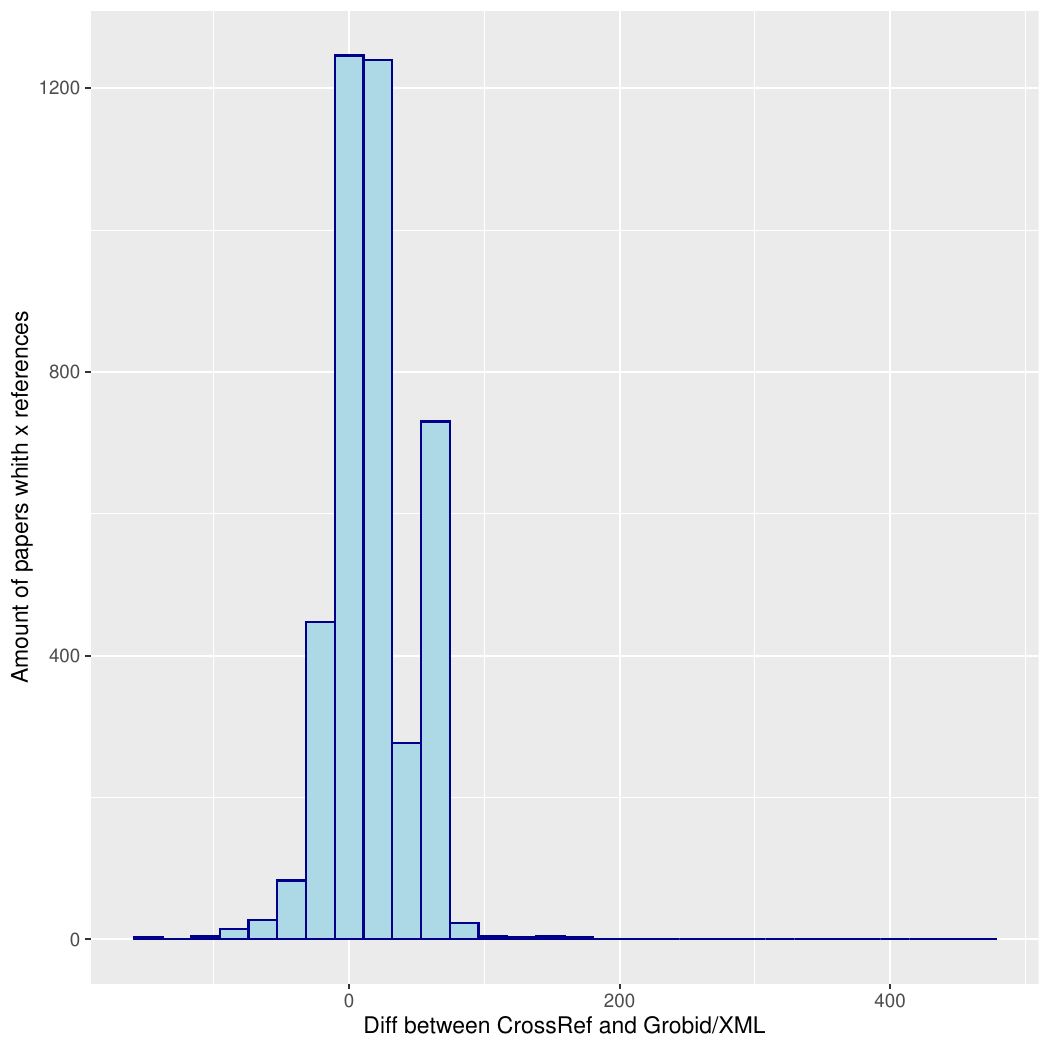}
		\subcaption{Distribution of the raw differences $\protect\mathcal R_{C}-\mathcal R_{G}$. Positive numbers are \SN, negative \LO}
		\label{fig:DiffBar}
	\end{subfigure}
\end{figure}

\end{document}